\documentclass[aps,prd,noshowpacs,preprintnumbers,nofootinbib,floatfix,twoside]{revtex4}

\usepackage{amsmath}
\usepackage{amssymb}
\usepackage{graphicx}
\usepackage{fancyhdr}
\usepackage{bm}

\newcommand{\Cal}[1]{\ensuremath{\mathcal{#1}}}
\newcommand{\dV}{\ensuremath{\partial\Cal{V}}}
\newcommand{\LL}{Lanczos-Lovelock }
\newcommand{\D}{\ensuremath{\nabla}}
\newcommand{\Riem}[4]{\ensuremath{R^{#1 #2}_{#3 #4}}}
\newcommand{\Alt}[6]{\ensuremath{\delta^{#1 #2 ... #3}_{#4 #5
      ... #6}}} 

\newcommand{\AltC}[8]{\ensuremath{\delta^{#1 #2 #3... #4}_{#5 #6 #7
      ... #8}}}  
\newcommand{\sD}[1]{\sum_{m=1}^{K}{#1}}
\newcommand{\LDm}{\ensuremath{\Cal{L}^{(D)}_m}}
\newcommand{\eqn}[1]{Eq.\eqref{#1}}
\newcommand{\ph}[1]{\phantom{#1}}

\newcommand{\om}{{\bm{\omega}}} %%%%%% Bold face \omega
\newcommand{\rmb}[1]{{\bf #1}} %%%%% Roman bold face
\newcommand{\ext}{{\rmb{d}}} %%%%%%%%% Exterior derivative
\newcommand{\msr}{{\bm{\epsilon}}}%% Natural D-form 
\newcommand{\wdot}{\,\dot\wedge\,} % ``Wedge-dot'' 

\begin{document}

\title{Entropy of Null Surfaces and Dynamics of Spacetime}

\author{T. Padmanabhan}
\email{paddy@iucaa.ernet.in}
\affiliation{IUCAA, Post Bag 4, Ganeshkhind, Pune - 411 007, India\\}

\author{Aseem Paranjape}
\email{aseem@tifr.res.in}
\affiliation{Tata Institute of Fundamental Research, Homi Bhabha
Road, Colaba, Mumbai - 400 005, India\\}

\date{\today}

%%%%%%%%%%%%%%%%%%%%%%%%%%%%%%%%%%%%%ABSTRACT%%%%%%%%%%%%%%%%%%%%%%%%%%%%%%%%%%%%%%%%
\begin{abstract}
The null surfaces of a spacetime act as one-way membranes and can
block information for a corresponding family of observers (time-like
curves). Since lack of information can be related to entropy, this
suggests the possibility of assigning an entropy to the null surfaces
of a spacetime. We motivate and introduce such an entropy functional for any vector field
in terms of  a fourth-rank
divergence free tensor $P_{ab}^{cd}$ with the symmetries of the
curvature tensor. Extremising this entropy for all the null surfaces then leads to equations for
the \textit{background metric} of the spacetime. When  $P_{ab}^{cd}$
is constructed from the metric alone, these equations are identical to
Einstein's equations with an undetermined cosmological constant (which
arises as an integration constant). More generally, if $P_{ab}^{cd}$
is allowed to depend on both metric and curvature in a polynomial
form, one recovers the \LL gravity. In all these cases: (a) We only
need to extremise the entropy associated with the null surfaces; the
metric is \textit{not} a dynamical variable in this approach. (b) The
extremal value of the entropy agrees with standard results, when
evaluated on-shell for a solution admitting a horizon. The role of
full quantum theory of gravity will be to provide the specific form of
$P_{ab}^{cd}$ which should be used in the entropy functional. With
such an interpretation, it seems reasonable to interpret the \LL type
terms as quantum corrections to classical gravity.

\end{abstract}

\maketitle
\vskip 0.5 in
\noindent
\maketitle

%%%%%%%%%%%%%%%%%%%%%%%%%%%%%%%%%%%%%%BODY%%%%%%%%%%%%%%%%%%%%%%%%%%%%%%%%%%%%%%%%%%%
\section{Introduction}
\noindent The strong mathematical resemblance between the dynamics of
spacetime horizons and thermodynamics has led several authors
\cite{elastic} to argue that a gravitational theory built upon the
Principle of Equivalence must be thought of as the \emph{macroscopic}
limit of some underlying microscopic theory. In particular, this
paradigm envisages gravity as analogous to the theory of elasticity of
a deformable solid. The unknown, microscopic degrees of freedom of
spacetime (which should be analogous to the atoms in the case of
solids), will play a role only when spacetime is probed at Planck
scales (which would be analogous to the lattice spacing of a solid
\cite{zeropoint}). Candidate models for quantum gravity, like e.g.,
string theory, do suggest the existence of such microscopic
degrees of freedom for gravity. The usual picture of treating the
metric as incorporating the dynamical degrees of freedom of the theory
is therefore not fundamental and the metric must be thought of as a
coarse grained description of the spacetime at macroscopic scales
(somewhat like the density of a solid which has no meaning at atomic  
scales).   

In such a picture, we expect the  microscopic structure of spacetime
to manifest itself only at Planck scales or near singularities of the
classical theory. However, in a manner which is not fully understood,
the horizons ---  which block information from certain classes of
observers --- link \cite{magglass} certain aspects of microscopic
physics with the bulk dynamics, just as thermodynamics can provide a
link between statistical mechanics and (zero temperature) dynamics of
a solid. (The  reason is probably related to the fact that horizons
lead to infinite redshift, which probes \textit{virtual} high energy
processes; it is, however, difficult to establish this claim in
mathematical terms). It has been known for several decades that one
can define the thermodynamic quantities entropy $S$ and temperature
$T$ for a spacetime horizon \cite{stdrefST}. If the above picture is
correct, then one should be able to link the equations describing bulk 
spacetime dynamics with horizon thermodynamics in a well defined
manner.   

There have been several recent approaches which have attempted to
quantify this idea with different levels of success
\cite{elastic,paddy1,paddyholo}. An explicit example  was
\cite{paddy2} the case of spherically symmetric horizons in four
dimensions. In this case, Einstein's equations can be interpreted as a 
thermodynamic relation $TdS=dE+PdV$ arising out of virtual
radial displacements of the horizon. More recently, it has been shown
\cite{aseem-sudipta} that this interpretation is not restricted to
Einstein's General Relativity (GR) alone, but is in fact true for the
case of the generalised, higher derivative \LL gravitational theory in
$D$ dimensions as well. Explicit demonstration of this result has also
been given for the case of Friedmann models in the \LL theory
\cite{rongencai} as well as for rotating and time dependent horizons
in Einstein's theory \cite{dawood-sudipta-tp}.  In a related development, there have been attempts to interpret other
gravitational Lagrangians (like $f(R)$ models) in terms of non-equilibrium thermodynamics \cite{cai2}.

In standard thermodynamics, extremisation of the functional form of
the entropy (treated as a function of the relevant dynamical
variables) leads to the equations governing the equilibrium state of
the system. This suggests that in the context of gravity as well, one
should be able to define an \emph{entropy functional}, which --- on
extremisation --- will lead to the equations describing the
macroscopic, long-wavelength behaviour of the system (which in this
case is the spacetime.) That is, if our analogy is to be taken
seriously, we should be able to define an ``entropy functional for
spacetime'', the extremisation of which should lead to the
gravitational field equations for the metric of the spacetime. At the
lowest order, this should give Einstein's equations but the formalism will
continue to be valid even in the quantum regime. It is important to
recall that --- even in the case of ordinary matter --- there is no
such thing as `quantum thermodynamics'; only quantum statistical
mechanics. What quantum theory does is to modify the form of the
entropy functional (or some other convenient thermodynamic potential,
like free energy); the extremisation now leads to equations of motion
which incorporate quantum corrections. Similarly, we expect our
entropy functional to pick up corrections to the lowest order term,
thereby leading to corrections to Einstein's equations. In that sense,
this approach is very general and this is what we will develop in this
paper. 

A crucial point to note is that the concept of entropy, both in the
standard thermodynamics as well as in the gravitational context, stems  
from the fact that certain degrees of freedom are \emph{not
  observable} for certain classes of observers. (Throughout the paper,
we will use the term observers to mean family of time like curves,
without any extra connotations.) In the context of a  
metric theory of gravity this is inevitably linked with the existence
of one-way membranes which are provided by \emph{null surfaces} in the
spacetime. The classical black hole event horizons, like e.g., the one
in the Schwarzschild spacetime, are special cases of such
one-way membranes. This also leads to the conclusion --- suggested by
several authors e.g., \cite{paddy1,obsdepentropy} ---
that the concept of entropy of spacetime horizons is intrinsically
observer dependent, since a null surface (one-way membrane) may act as
a horizon for a certain class of observers but not for some other
class of observers. In flat Minkowski spacetime, the light cone at
some event can act as a horizon for the appropriate class of uniformly
accelerated Rindler observers, but not so for inertial
observers. Similarly, even the black hole horizon (which can be given
a `geometrical' definition, say, in terms of a Penrose diagram) will be
viewed differently by an observer falling into the black hole compared
to another who is  orbiting at a radius $r>2M$. The fact that the
dynamics of the spacetime should nevertheless be described in an
observer independent manner leads to very interesting consequences, as
we shall see. 

The above discussion also points to the possibility that the 
 null surfaces of the spacetime should play a key role in the extremum principle we want to develop.
 It is also important, within this conceptual
framework, that the metric is not a dynamical variable but an emergent,
long wavelength concept \cite{comment0}.  In this paper, we will
construct such an entropy functional in  a
metric theory of gravity and derive the equations resulting from its
extremisation. We will show that not only Einstein's GR, but also the
higher derivative \LL theory can be naturally incorporated in this
framework. The formalism also allows us to write down higher order
quantum corrections to Einstein's theory in a systematic, algorithmic
procedure.

The paper is organized as follows: In section 2 we motivate
a definition for the entropy functional $S$ related to a
vector field $\xi^a$  in the spacetime  in the context
of the Einstein and \LL theories, and elaborate upon the variational
principle we employ to determine the spacetime dynamics from this functional. In
section 3 we compute the extremised value of $S[\xi]$ and show that
under appropriate circumstances it is identical to the expression for
the horizon entropy as derived by other authors in the context of \LL
theories, thus justifying (at least partially) the name `entropy functional'. 
Section 4 rephrases our results in the language of forms, to give its
geometrical meaning. We conclude in section 5 by discussing some
implications of our results.

\section{An Entropy Functional for Gravity}

Our key task is to define a suitable entropy functional for the
spacetime. Since this is similar to introducing the action functional
for the theory, it is obvious that we will not be able to
\textit{derive} its form without knowing the microscopic theory. So we
shall do the next best thing of motivating its choice. (If the reader
is unhappy with the motivating arguments, (s)he may take the final form
of the entropy functional in \eqn{ent-func-2} below as the basic
postulate of our approach!) 

The first clue comes from the theory of elasticity. We know that, in
the theory of elasticity \cite{lan-lif},  the key quantity is a \textit{displacement vector field} $\xi^a(x)$
which describes the elastic displacement of the solid through the equation $x^a\to x^a+\xi^a(x)$. (Of course, in elasticity, one usually deals with three-vectors while we need to work in D-dimensions!). All thermodynamic potentials, including
the entropy of a deformable solid can be written as an integral over a
quadratic functional of the displacement vector field, which can
capture the relevant dynamics in the long-wavelength limit. In the
context of gravity, the ``solid'' in question is spacetime itself
\cite{comment1}. The crucial difference from the theory of elasticity
is the following: In elasticity, extremising the entropy function will
lead to an equation \textit{for} the displacement field and determine
$\xi^a$. In the case of spacetime, the equations should determine  the \textit{background
metric.} This is a nontrivial constraint on the structure of
the theory and we will show how this can be achieved.  

In the case of elasticity as well as gravity, we will expect the
entropy functional to be an integral over a local entropy density, so
that extensivity on the volume is ensured. In the case of an elastic
solid, we expect the entropy density to be translationally invariant
and hence depend only upon the derivatives of $\xi^a$ quadratically to
the lowest order. We would expect this to be true for \textit{pure}
gravity as well, and hence the entropy density should have a form
$P_{ab}^{\ph{a}\ph{b}cd} \D_c\xi^a\D_d\xi^b$, where the fourth
rank (tensorial) object $P_{ab}^{\ph{a}\ph{b}cd}$ is built out of
metric and other geometrical quantities like curvature tensor of
the background spacetime. But in the presence of non-gravitational
matter distribution in spacetime (which, alas, has no geometric
interpretation), one cannot demand translational invariance. Hence,
the entropy density can have quadratic terms in both the derivatives
$\D_a\xi^b$ as well as $\xi^a$ itself. We will denote the latter
contribution as $T_{ab}\xi^a\xi^b$ where the second rank tensor
$T_{ab}$ (which is taken to be symmetric, since only the symmetric
part is relevant to this expression) is determined by matter
distribution and will vanish in the absence of matter. (We will later 
see that $T_{ab}$ is just the energy momentum tensor of matter; the
notation anticipates this but does not demand it at this stage.) So
our entropy functional can be written as:   
\begin{equation}
S[\xi]=\int_\Cal{V}{d^Dx\sqrt{-g}}
    \left(4P_{ab}^{\ph{a}\ph{b}cd} \D_c\xi^a\D_d\xi^b - 
    T_{ab}\xi^a\xi^b\right) \,,
\label{ent-func-2}
\end{equation}
where \Cal{V}\ is a $D$-dimensional region in the spacetime with
boundary \dV, and we have introduced some additional factors and signs
in the expression for later convenience. We will now impose two
additional conditions on $P_{ab}^{\ph{a}\ph{b}cd}$ and $T_{ab}$. 
(i) For the case of the elastic solid, the coefficients of the
quadratic terms are constants (related to to the bulk modulus, the
modulus of rigidity and so on). We take the analogues of these
constant coefficients to be quantities with vanishing covariant
divergences. That is, we postulate the ``constancy'' conditions:  
\begin{equation}
\D_{b}P_{a}^{\ph{a}bcd}=0=\D_{a}T^{ab}\,.
\label{ent-func-1}
\end{equation}
(ii) The second  requirement we impose is that the tensor $P_{abcd}$ should
have the algebraic symmetries similar to the Riemann tensor $R_{abcd}$
of the $D$-dimensional spacetime; viz.,  $P_{abcd}$ is
antisymmetric in $ab$ and $cd$ and symmetric under pair exchange. 
\eqn{ent-func-1} then implies that the $P^{abcd}$ will be divergence-free in
\emph{all} its indices. Because of these symmetries, the notation
$P^{ab}_{cd}$ with two upper and two lower indices is unambiguous.   

In summary, we associate with every  vector field $\xi^a$ in the
spacetime an entropy functional in \eqn{ent-func-2}, with the
conditions: (i) The tensor $P_a^{\ph{a}bcd}$ is built from background
geometrical variables, like the metric, curvature tensor, etc. and has
the algebraic index symmetries of the curvature tensor. It is also
divergence free. (ii) The tensor $T_{ab}$ is related to the matter
variables and vanishes in the absence of matter. It also has zero
divergence.
One key feature of the functional in \eqn{ent-func-2} is
that the entropy associated with \textit{null} vector fields are invariant under the shift $T_{ab}\to T_{ab}+\rho g_{ab}$
where $\rho$ is a scalar. This fact will play an interesting role later on.  

\subsubsection{Explicit form of $P^{abcd}$}

Obviously, the structure of the gravitational sector is encoded in the
form of $P^{abcd}$ and we need to consider the possible choices for
$P^{abcd}$ which determine the form of the entropy functional. In a
complete theory, the form of $P^{abcd}$ will be determined by the
long wavelength limit of the microscopic theory just as the elastic
constants can --- in principle --- be determined from the microscopic
theory of the lattice. However, our situation in gravity is similar to
that of the physicists of the eighteenth century with respect to
solids and --- just like them ---  we need to determine the ``elastic 
constants'' of spacetime by general considerations. Taking a cue from
the standard approaches in renormalization group, we expect $P^{abcd}$
to have a derivative expansion in powers of number of derivatives of
the metric:  
\begin{equation}
P^{abcd} (g_{ij},R_{ijkl}) = c_1\,\overset{(1)}{P}{}^{abcd} (g_{ij}) +
c_2\, \overset{(2)}{P}{}^{abcd} (g_{ij},R_{ijkl})  
+ \cdots \,,
\label{derexp}
\end{equation} 
where $c_1, c_2, \cdots$ are coupling constants.  The lowest order
term must clearly depend only on the metric with no derivatives. The next
term depends on the metric and curvature tensor. Note that since
$P^{abcd}$ is a tensor, its expansion in derivatives of the metric
necessarily involves the curvature tensor as a ``package'' comprising
of products of first derivatives of the metric (the $\Gamma\Gamma$
terms) and terms linear in the second derivatives ($\partial\Gamma$),
where $\Gamma$ symbolically denotes the Christoffel connection. Higher
order terms can involve both higher powers of the curvature tensor, as
well as its covariant derivatives. 

These terms can, in fact, be listed from the required symmetries of
$P_{abcd}$. For example, let us consider the possible fourth rank
tensors $P^{abcd}$ which (i) have the symmetries of curvature tensor;
(ii) are divergence-free; (iii) are made from $g^{ab}$ and
$R^a_{\ph{a}bcd}$ but not derivatives of $R^a_{\ph{a}bcd}$. If we do
not use the curvature tensor, then we have just one choice 
made from the metric:
\begin{equation}
\overset{(1)}{P}{}^{ab}_{cd}=\frac{1}{32\pi}
(\delta^a_c \delta^b_d-\delta^a_d \delta^b_c)  \,.
\label{pforeh}
\end{equation} 
We have fixed an arbitrary constant in the above expression for later
convenience. Next, if we allow for $P^{abcd}$ to depend
linearly on curvature, then we have the following additional choice of
tensor with the required symmetries: 
\begin{equation}
\overset{(2)}{P}{}^{ab}_{cd}=\frac{1}{8\pi} \left(R^{ab}_{cd} -
         G^a_c\delta^b_d+ G^b_c \delta^a_d +  R^a_d \delta^b_c -
         R^b_d \delta^a_c\right)   \,.
\label{ping}
\end{equation} 
We have again chosen an arbitrary constant for convenience, but in
this case the constant can always be specified in the factor
$c_2$ of \eqn{pforeh}. 

The expressions in \eqn{pforeh} and \eqn{ping} can be expressed in a
 more illuminating form. Note that, the expression 
 in \eqn{pforeh} is just
\begin{equation}
\overset{(1)}{P}{}^{a_1a_2}_{b_1b_2} = \frac{1}{16\pi}
\frac{1}{2} \delta^{a_1a_2}_{b_1b_2} \,,
\label{alt2}
\end{equation}
where we have the introduced the alternating or `determinant' tensor
 $\delta^{a_1a_2}_{b_1b_2}$. Similarly, the expression in \eqn{ping}
 above can be rewritten in the following form: 
\begin{equation}
\overset{(2)}{P}{}^{a_1a_2}_{b_1b_2} = \frac{1}{16\pi}
\frac{1}{2} \delta^{a_1a_2a_3a_4}_{b_1\,b_2\,b_3\,b_4}
R^{b_3b_4}_{a_3a_4}  \,.  
\label{alt1}
\end{equation}
where we have again introduced the alternating  tensor
$\delta^{a_1a_2a_3a_4}_{b_1\,b_2\,b_3\,b_4}$ 
\begin{equation}
\delta^{a_1a_2a_3a_4}_{b_1\,b_2\,b_3\,b_4} = \frac{-1}{(D-4)!}
  \epsilon^{c_1\cdots c_{D-4}a_1a_2a_3a_4}\epsilon_{c_1\cdots
  c_{D-4}b_1b_2b_3b_4} 
  \,, 
\label{alteps}
\end{equation}
The alternating tensors are totally antisymmetric in both sets of
indices and take values $+1$, $-1$ and $0$. They can be written in any
dimension as an appropriate contraction of the Levi-Civita tensor
density with itself \cite{MTW}. (In 4 dimensions the expression  in
\eqn{ping} is essentially the double-dual of $R_{abcd}$.)   
We see a clear pattern emerging from \eqn{alt2} and
\eqn{alt1} with the $m$-th order contribution being a term involving
$(m-1)$ factors of the curvature tensor. Following this pattern it is
easy to construct the $m$-th order term 
which satisfies our constraints. This is unique and is given by
\begin{equation}
\overset{(m)}{P}{}_{ab}^{\ph{a}\ph{b}cd}\propto
\AltC{c}{d}{a_3}{a_{2m}}{a}{b}{b_3}{b_{2m}}
\Riem{b_3}{b_4}{a_3}{a_3} \cdots
\Riem{b_{2m-1}}{b_{2m}}{a_{2m-1}}{a_{2m}}  
\,.
\label{LL03}
\end{equation}
These terms have a close relationship with the Lagrangian for  \LL
theory, which is a generalised higher derivative theory of gravity. Before proceeding further, we 
shall briefly recall the properties of \LL theory and describe this
connection. 

The \LL Lagrangian is a specific example from a general
class of Lagrangians which describe a (possibly semiclassical) theory
of gravity and are given by    
\begin{equation}
\Cal{L}=Q_a^{\ph{a}bcd}R^a_{\ph{a}bcd}\,,
\label{LL1}
\end{equation}
where $Q_a^{\ph{a}bcd}$ is the most general fourth rank tensor sharing
the algebraic symmetries of the Riemann tensor $R^a_{\ph{a}bcd}$ and further
satisfying the criterion $\D_bQ_a^{\ph{a}bcd}=0$ (Several general properties of this class of Lagrangians are discussed in Ref. \cite{ayan}). The $D$-dimensional
\LL Lagrangian is given by \cite{lovelock} a polynomial in the
curvature tensor: 
\begin{equation}
\Cal{L}^{(D)} = \sD{c_m\LDm}\,~;~\Cal{L}^{(D)}_m = \frac{1}{16\pi}
2^{-m} \Alt{a_1}{a_2}{a_{2m}}{b_1}{b_2}{b_{2m}}
\Riem{b_1}{b_2}{a_1}{a_2} \Riem{b_{2m-1}}{b_{2m}}{a_{2m-1}}{a_{2m}}
\,,  
\label{LL2}
\end{equation}
where the $c_m$ are arbitrary constants and \LDm\ is the $m$-th
order \LL term. Here the generalised alternating 
tensor $\delta^{\cdots}_{\cdots}$  is the natural extension of the one
defined in \eqn{alteps} for $2m$ indices, and we assume
$D\geq2K+1$. The $m$-th order \LL term $\Cal{L}^{(D)}_m$ given  
in \eqn{LL2} is a homogeneous function of the Riemann tensor of degree
$m$. For each such term, the tensor $Q_a^{\ph{a}bcd}$ defined in
\eqn{LL1} carries a label $m$ and becomes 
\begin{equation}
{}^{(m)}Q_{ab}^{\ph{a}\ph{b}cd}=  \frac{1}{16\pi}2^{-m}
\AltC{c}{d}{a_3}{a_{2m}}{a}{b}{b_3}{b_{2m}}
\Riem{b_3}{b_4}{a_3}{a_3}\cdots\Riem{b_{2m-1}}{b_{2m}}{a_{2m-1}}{a_{2m}}
\,.
\label{LL3}
\end{equation}
The full tensor $Q_{ab}^{\ph{a}\ph{b}cd}$ is a linear combination of
the ${}^{(m)}Q_{ab}^{\ph{a}\ph{b}cd}$ with the coefficients $c_m$.
Einstein's GR is a special case of \LL gravity in which only the
coefficient $c_1$ is non-zero. Since the tensors
${}^{(m)}Q_{ab}^{\ph{a}\ph{b}cd}$ appear linearly in the \LL
Lagrangian and consequently in all other tensors constructed from it,
it is sufficient to concentrate on the case where a single coefficient
$c_m$ is non-zero. All the results that follow can be easily extended
to the case where more than one of the $c_m$ are non-zero, by taking
suitable linear combinations of the tensors involved. Hence, to avoid
displaying cumbersome notation and summations,  we will now restrict our
attention to a single $m$-th order \LL term \LDm, and will also
drop the superscript $(m)$ on the various quantities. 
Comparing with our expression in \eqn{LL03} it is clear that
$P_a^{\ph{a}ijk}$ can be taken to be proportional to $Q_a^{\ph{a}ijk}$
which  can be conveniently expressed
as a derivative of the \LL Lagrangian with respect to the curvature
tensor. To be concrete, we shall take the  $m$-th order term in
\eqn{LL03} to be:  
\begin{equation}
P_a^{\ph{a}ijk}= mQ_a^{\ph{a}ijk}=M_a^{\ph{a}ijk} \equiv
\frac{\partial\LDm}{\partial R^a_{{\ph{a}ijk}}}\,. 
\label{LL8}
\end{equation}
This equation defines the divergence-free tensor $M_a^{\ph{a}ijk}$,
where the partial derivatives are taken treating $g^{ab}$,
$\Gamma^a_{\ph{a}bc}$ and $R^a_{\ph{a}bcd}$ as independent quantities.  
The numerical coefficients are chosen for later convenience and can be, of course,
absorbed into the definitions of the $c_m$. With this choice, we have
completely defined the geometrical structure of the entropy
functional, except for the coupling constants $c_m$ which appear at
each order \cite{comment11}. 

Just to see explicitly and in gory detail what we have, let us write 
down the entropy functional in the absence of matter ($T_{ab}=0$),
correct up to  first order in the curvature tensor in 
$P^{abcd}$. To this order, our entropy functional \eqref{ent-func-2}
takes the form $S=S_1+S_2$ where
\begin{eqnarray}
S_1[\xi]&=&\int_\Cal{V}\frac{d^Dx}{8\pi}
   \left((\D_c\xi^c)^2-\D_a\xi^b\D_b\xi^a\right)\nonumber\\
S_2[\xi]&=&c_2\int_\Cal{V}d^Dx
  \left( R_{ab}^{cd}\D_c\xi^a\D_d\xi^b -
    (G^c_a+R^c_a)(\D_c\xi^a\D_b\xi^b - \D_c\xi^b\D_b\xi^a) \right)
\label{2order}
\end{eqnarray}
where $c_2$ is a coupling constant and we have used equations
\eqref{pforeh} and \eqref{ping}. 
We will later see that the entropy given by $S_1$ leads to Einstein's
equations in general relativity while $S_2$ and higher order terms can
be interpreted as corrections to this. 
If we choose $\xi^a$ to be the normal vector field of a sequence of hypersurfaces foliating the spacetime, then the
integrand in $S_1$ has the
familiar structure, $ (TrK)^2-Tr(K^2)$ where $K_{ab}$
is the extrinsic curvature. (This could offer an alternative interpretation of ADM action; however, we will not discuss this aspect here except to cast our results in the familiar language when appropriate.)
The expression in \eqn{2order} can be further simplified by
integrating it by parts where appropriate and writing the right hand
side as a sum of a contribution from the bulk and a surface term. The
general expression after such a splitting is given later in
\eqn{on-shell-1} and can also be found in Section 4 where it arises
transparently in the language of forms (see \eqn{form-19}).  

\subsubsection{Field equations from extremising the entropy}

Having made these general observations regarding the choice of
$P^{ab}_{cd}$ let us now return to the entropy functional in
\eqn{ent-func-2}. This expression is well defined for \textit{any}
displacement vector field $\xi^a$. We can, therefore, associate an
entropy functional with any hypersurface in the spacetime, by choosing
the normal to the hypersurface as $\xi^a$. Among all such
hypersurfaces, the null hypersurfaces will play a key role since they
act as one-way membranes which block information for a specific class
of observers. Given this motivation, we  will now extremise this $S$ with respect to
variations of the null vector field $\xi^a$ and demand that the
resulting condition holds for \textit{all null vector fields}. That is, the
``equilibrium'' configurations of the ``spacetime solid'' are the ones
in which the entropy associated with \textit{every} null vector is
extremised. Varying the null vector field $\xi^a$ after adding a
Lagrange multiplier $\lambda$ for imposing the null condition
$\xi_a\delta \xi^a=0$, we find: 
\begin{eqnarray}
\delta S &=& 2\int_\Cal{V}{d^Dx\sqrt{-g}
  \left(4P_{ab}^{\ph{a}\ph{b}cd}\D_c\xi^a\left(\D_d\delta\xi^b\right)
  - T_{ab}\xi^a\delta\xi^b - \lambda g_{ab}\xi^a\delta\xi^b\right)}
  \nonumber\\ 
  &\equiv& 2\int_\Cal{V}{d^Dx\sqrt{-g}
  \left(4P_{ab}^{\ph{a}\ph{b}cd}\D_c\xi^a\left(\D_d\delta\xi^b\right)
  - \bar T_{ab}\xi^a\delta\xi^b \right)} 
\label{ent-func-3}
\end{eqnarray}
where we have used the symmetries of $P_{ab}^{\ph{a}\ph{b}cd}$ and
$T_{ab}$ and set $\bar T_{ab}=T_{ab}+\lambda g_{ab}$. (As we said
before such a shift leaves entropy associated with null vectors unchanged so the Lagrange
multiplier will turn out to be irrelevant; nevertheless, we will use
$\bar T_{ab}$ for the moment.) An integration by parts and the
condition $\D_dP_{ab}^{\ph{a}\ph{b}cd}=0$, leads to 
\begin{equation}
\delta
S=2\int_\Cal{V}{d^Dx\sqrt{-g}\left(-4P_{ab}^{\ph{a}\ph{b}cd}
  \left(\D_d\D_c\xi^a\right) - \bar T_{ab}\xi^a\right)\delta\xi^b} +
8\int_{\dV}{d^{D-1}x\sqrt{h}\left(n_d
  P_{ab}^{\ph{a}\ph{b}cd}\left(\D_c\xi^a\right)\right)\delta\xi^b}
\,,
\label{ent-func-4}
\end{equation}
where $n^a$ is the $D$-vector field normal to the boundary \dV\ and
$h$ is the determinant of the intrinsic metric on \dV.  As usual, in order for
the variational principle to be well defined, we require that the
variation $\delta\xi^a$ of the null vector field should vanish on the
boundary. The second term in \eqn{ent-func-4} therefore vanishes, and
the condition that $S[\xi]$ be an extremum for arbitrary variations of
$\xi^a$ then becomes  
\begin{equation}
2P_{ab}^{\ph{a}\ph{b}cd}\left(\D_c\D_d-\D_d\D_c\right)\xi^a
-\bar T_{ab}\xi^a = 0\,,
\label{ent-func-5}
\end{equation}
where we used the antisymmetry of $P_{ab}^{\ph{a}\ph{b}cd}$ in its
upper two indices to write the first term. The definition of the
Riemann tensor in terms of the commutator of covariant derivatives
reduces the above expression to
\begin{equation}
\left(2P_b^{\ph{b}ijk}R^a_{\ph{a}ijk} - \bar T{}^a_b\right)\xi_a=0\,, 
\label{ent-func-6}
\end{equation}
and we see that the equations of motion \emph{do not contain}
derivatives with respect to $\xi$. This peculiar feature arose because
of the symmetry requirements we imposed on the tensor
$P_{ab}^{\ph{a}\ph{b}cd}$. We further require that the condition in
\eqn{ent-func-6} hold for \emph{arbitrary} null vector fields
$\xi^a$. A simple argument based on local Lorentz invariance then
implies that 
\begin{equation}
2P_{b}^{\ph{b}ijk}R^{a}_{\ph{a}ijk} - T{}_b^a = F(g)\delta{}_b^a\,,   
\label{ent-func-7}
\end{equation}
where $F(g)$ is some scalar functional of the metric and we have
absorbed the $\lambda\delta^a_b$ in $\bar T{}^a_b =
T{}^a_b+\lambda\delta^a_b$ into the definition of $F$.  The validity 
of the result in \eqn{ent-func-7} is obvious if we take a dot product
of \eqn{ent-func-6} with $\xi^b$. (A formal proof can be found in the
Appendix\eqref{sec:loc-Lor}.) The scalar $F(g)$ is arbitrary so far
and we will now show how it can be determined in the physically
interesting cases.  

\medskip
\centerline{\textit{2.1 Lowest order theory: Einstein's equations}}
\medskip
\noindent To do this, let us substitute the derivative expansion for
$P^{abcd}$ in \eqn{derexp} into  \eqn{ent-func-7}. To the lowest order
we find that the equation reduces to: 
\begin{equation}
\frac{1}{8\pi}R^a_b - T^a_b=F(g)\delta^a_b 
\end{equation}
where $F$ is an arbitrary function of the metric. Writing this
equation as $(G^a_b - 8\pi T^a_b ) = Q(g) \delta^a_b $ with $Q= 8\pi
F- (1/2) R$ and using $\nabla_a G^a_b = 0 , \nabla_a T^a_b =0$ we get 
$\partial_bQ=\partial_b [ 8\pi F - (1/2) R] =0$; so that $Q$ is an
undetermined integration constant, say $\Lambda$, and $F$ must have
the form  $8\pi F=(1/2)R+\Lambda$. The resulting equation is
\begin{equation}
R^a_b-(1/2)R\delta^a_b=8\pi T^a_b+\Lambda\delta^a_b
\label{eom}
\end{equation}
which leads to Einstein's theory  if we identify $T_{ab}$ as the
matter energy momentum tensor \textit{with a cosmological constant
  appearing as an integration constant}. (For the importance of the
latter with respect to the cosmological constant problem, see
Ref. \cite{cc1}; we will not discuss this issue here.) 

The same procedure works with the first order term in \eqn{derexp} as 
well and we reproduce the Gauss-Bonnet gravity with a cosmological
constant. In this sense, we can interpret the first term in the
entropy functional in \eqn{2order} as the entropy in Einstein's
general relativity and the term proportional to $c_2$ as a
Gauss-Bonnet correction term. 
Instead of carrying out this analysis explicitly order by order, we shall now describe
the most general structure in the family of theories starting with 
Einstein's GR, Gauss-Bonnet gravity etc. ---  and will show that we
reproduce the \LL theory by our approach. 

\medskip
\centerline{\textit{2.2 Higher order corrections: \LL gravity}}
\medskip
\noindent 
To see this result, let us briefly recall some aspects of \LL theory. It can
be shown  that (see e.g., \cite{ayan}) the equations of motion 
for a general theory of gravity derived from the Lagrangian in
\eqn{LL1} using the standard variational principle with $g^{ab}$ as
the dynamical variables, are given by  
\begin{equation}
E_{ab}=\frac{1}{2}T_{ab} ~~;~~
  E_{ab}\equiv\frac{1}{\sqrt{-g}}\frac{\partial}{\partial
  g^{ab}}\left(\sqrt{-g}\Cal{L}\right) -2\D^m\D^nM_{amnb}\,. 
\label{LL4}
\end{equation}
Here $T_{ab}$ is the energy-momentum tensor for the matter fields. The 
tensor $M_{abcd}$ defined through
$M_a^{\ph{a}bcd}\equiv(\partial\Cal{L}/\partial R^a_{\ph{a}bcd})$ is a
generalisation of the one defined for the \LL case in \eqn{LL8}.
The partial derivatives are as before taken treating $g^{ab}$, 
$\Gamma^a_{\ph{a}bc}$ and $R^a_{\ph{a}bcd}$ as independent
quantities. For the $m$-th order \LL Lagrangian $\LDm$, since
$M^{abcd}$ is divergence-free, the expression for the tensor $E_{ab}$
in \eqn{LL4} becomes 
\begin{equation}
E_{ab}=\frac{\partial\LDm}{\partial g^{ab}}-\frac{1}{2}\LDm 
g_{ab}\,, 
\label{LL5}
\end{equation}
where we have used the relation $\partial(\sqrt{-g})/\partial
g^{ab}=-(1/2)\sqrt{-g}g_{ab}$. The first term in the expression for
$E_{ab}$ in \eqn{LL5} can be simplified to give 
\begin{equation}
\frac{\partial\LDm}{\partial
  g^{ab}}=mQ_{a}^{\ph{a}ijk}R_{bijk}= M_a^{\ph{a}ijk}R_{bijk} \,,  
\label{LL6}
\end{equation}
where the expressions in \eqn{LL6} can be verified by direct
computation, or by noting that \LDm\ is a homogeneous function 
of the Riemann tensor $R^a_{\ph{a}bcd}$ of degree $m$. To summarize,
the \LL field equations are given by
\begin{equation}
E_{ab}=\frac{1}{2}T_{ab} ~~;~~
   E_{ab}= mQ_{a}^{\ph{a}ijk}R_{bijk} -\frac{1}{2}\LDm 
   g_{ab} \,. 
\label{LL7}
\end{equation}
Further, diffeomorphism invariance implies that the tensor $E_{ab}$
defined in \eqn{LL4} is divergence-free, $\D_aE{}^a_b=0$. The
equations of motion for the matter imply that the energy-momentum
tensor $T_{ab}$ is also divergence-free (as required by
\eqn{ent-func-1}). Using these conditions in 
\eqn{ent-func-7} together with the choice in \eqn{LL8} for
$P_a^{\ph{a}ijk}$ leads to
\begin{equation}
\partial_aF=\partial_a\LDm\,,
\label{LL9}
\end{equation}
which fixes $F(g)$ as $F=\LDm+\Lambda/8\pi$ where $\Lambda$ is a
constant with the normalisation chosen so as to conform with the usual
definition of the cosmological constant. The resulting field equations for \LL gravity will be:
\begin{equation}
16\pi\left[ P_{b}^{\ph{b}ijk}R^{a}_{\ph{a}ijk}-\frac{1}{2}\delta^a_b\LDm\right]=
 8\pi T{}_b^a,   
\label{ent-func-71}
\end{equation}
where we have included a possible cosmological constant, that arises as an undetermined integration constant in the defintion of $T^a_b$. Taking the trace of this equation, we find that that $\LDm=(2m-D)^{-1}T$. In other words, the on-shell value of the Lagrangian is proportional to the trace of the stress tensor in all \LL theories, just like in GR. In the absence of source term, this implies that $\LDm=0$ and
the equations of motion reduces to $P_{b}^{\ph{b}ijk}R^{a}_{\ph{a}ijk}=0$.
The case $m=1$ with $\Cal{L}^{(D)}_{m=1} = (1/16\pi)R$ is easily
seen to reduce to that of Einstein's gravity.
.

To summarise, if we take the derivative expansion in \eqn{derexp} to
correspond to a polynomial form in the curvature tensor, then it has
the form given by \eqn{LL8}. In this case,
extremising the entropy leads to the \LL theory. We stress that the
resulting field equations have the form of Einstein's equations with
higher order corrections. In our picture, we consider this as emerging  
from the form of the entropy functional which has an expansion in
powers of the curvature. 

Before concluding this section, we want  to comment on an interesting
property of the entropy functional. The derivation of the equations
\eqref{ent-func-7} was based upon a variational principle which 
closely resembles the usual variational principle used in other areas of physics in which some quantity is varied within an integral arbitrarily, except for it being fixed at the boundary. Instead of such an arbitrary variation, let us consider a \emph{subset} of all possible  
variations of the null vector field $\xi^a$, given by
$\xi^a(x)\to(1+\epsilon(x))\xi^a(x)$; namely infinitesimal
\emph{rescalings} of $\xi^a$. We assume that the scalar $\epsilon(x)$
is infinitesimal and also that it vanishes on the boundary \dV. In
this case it is easy to see that the variation of $S[\xi]$ in
\eqn{ent-func-4} becomes 
\begin{equation}
\delta S|_{\rm rescale}=2\int_\Cal{V}{d^Dx\sqrt{-g}\left(
  2P_b^{\ph{b}ijk}R^a_{\ph{a}ijk} -
  T{}^a_b\right)\xi_a\xi^b\epsilon(x)}
\label{invar1}
\end{equation}
Clearly, requiring that the functional $S[\xi]$ be \emph{invariant}
under rescaling transformations of $\xi^a$ leads to the same
requirement as before, namely that \eqn{ent-func-7} be satisfied.
We can understand the physical motivation behind imposing such a
symmetry condition on $S[\xi]$ as follows. Let us begin by noting the
fact that the causal structure of a spacetime, which can be thought of
as the totality of all possible families of null hypersurfaces in the
spacetime, is left invariant under rescalings of the generators of
these null hypersurfaces. To see this symmetry, note that \emph{any}
curve in the spacetime, with tangent vector field $t^a=dx^a/d\lambda$, say, is
invariant under the rescaling $t^a\to f(\lambda)t^a$ of the tangent field,
where $f$ is some scalar. This is so because a rescaling of the
tangent vector field is equivalent to a reparametrization of the
curve. A null hypersurface \Cal{H}\ can be thought of as being 
`filled'  by null geodesics contained in it, and is hence invariant
under rescalings of its generator field $\xi^a$. The result for the
full causal structure then follows. The functional $S[\xi]$ depends
only on the generator $\xi^a$ of some null hypersurface (apart from
the metric and matter fields which we consider as given
quantities). It is therefore natural to demand that this symmetry
of the causal structure also be a symmetry of $S[\xi]$.  We do not, however, use this feature in this paper. (For
a completely different, purely classical, approach to general
relativity based on null surfaces, see \cite{othernull}).

\section{Evaluating the Entropy Functional On-Shell}

The results in the previous section show that our approach provides -- at the least -- an alternative variational principle to obtain not only  Einstein's theory but also \LL theory. In the conventional approaches to these theories, one can obtain the field equations by varying
the action functional. But once the field equations are obtained, the extremum \textit{value} of the action functional is not of much concern.
(The only exceptions are in the semiclassical limit in which it appears as the phase of the wave function or in some specific Euclidean extension of the solution.) In our approach, it is worthwhile to proceed further and ask what this extremal  [`on-shell'] value means in specific contexts.
We will be able to provide an interpretation under some specific contexts, justifying the term entropy functional but it should be stressed that the results in this section are logically independent of the derivation of field equations in the previous section. In particular, we need to consider non-null vectors to provide a natural interpretation of the extremum value of the functional.

The term `on-shell' refers to satisfying the relevant equations of
motion, which in this case are given by \eqn{ent-func-6}. Manipulating
the covariant derivatives in \eqn{ent-func-2}, we can write 
\begin{align}
S[\xi]&=\int_\Cal{V}{d^Dx\sqrt{-g}\left[
    4\D_d\left(P_{ab}^{\ph{a}\ph{b}cd}\left(
    \D_c\xi^a\right)\xi^b\right) 
    - 4P_{ab}^{\ph{a}\ph{b}cd}\left(\D_d\D_c\xi^a\right)\xi^b
    -T_{ab}\xi^a\xi^b \right]} 
\nonumber\\
&=4\int_{\dV}{d^{D-1}x\sqrt{h}
    n_d\left(P_{ab}^{\ph{a}\ph{b}cd}\xi^b\D_c\xi^a\right)}
    +  \int_\Cal{V}{d^Dx\sqrt{-g}\left(
    2P_{mb}^{\ph{a}\ph{b}cd}R^m_{\ph{m}acd} 
    - T_{ab}\right)\xi^a\xi^b} \,.
\label{on-shell-1}
\end{align}
In writing the first equality, we have used the condition
$\D_dP_{a}^{\ph{a}bcd}=0$. As before, in the first term of second
equality, $n^a$ is the vector field normal to the boundary \dV\ and
$h$ is the determinant of the intrinsic metric on \dV. (In general, the boundary is 
$(D-1)$-dimensional. We will soon see that the really interesting case
occurs, in fact, when  part of the boundary \dV\ is
null and hence intrinsically $(D-2)$-dimensional. This case needs to be
handled by a limiting procedure and in what follows we will elaborate on the
procedure we use.) The second term of the second equality in
\eqn{on-shell-1}  vanishes in the absence of matter because, when $T_{ab}=0$, the equations of motion reduces to $Q_{a}^{\ph{a}ijk}R_{bijk}=0$, thereby allowing us to interpret the first term as the on-shell value of
  entropy from the gravity sector. Even in the presence of matter,  the second term can be expressed in terms of matter variables as an integral over the trace of the stress tensor (see the discussion around \eqn{ent-func-71}) and, of course, is not a surface term. We will, therefore, concentrate on the surface term arising from the gravitational sector which reduces to
\begin{equation}
S|_{\rm on-shell}=4\int_{\dV}{d^{D-1}x\sqrt{h}\,n_a\left(P^{abcd}\xi_c\D_b\xi_d\right)}
\longrightarrow\frac{1}{8\pi}\int_{\dV}d^{D-1}x\sqrt{h}\,n_a\left(\xi^a\D_b\xi^b-\xi^b\D_b\xi^a\right)
\label{on-shell-2}
\end{equation}
where we have manipulated a few indices using the symmetries of
$P^{abcd}$. The second expression after the arrow is the result for
general relativity; we give this explicitly to show the form of the
expression in a familiar setting. Note that, when $\xi^a$ is chosen as the normal to a set of surfaces foliating the spacetime, the integrand has the
familiar structure of $n_i(\xi^iK+a^i)$ where $a^i=\xi^b\D_b\xi^i$ is
the acceleration associated with the vector field $\xi^a$ and
$K\equiv -\D_b\xi^b$ is the trace of extrinsic
curvature in the standard context. This is the standard surface term which arises in the ADM formulation and the cognoscenti will immediately see its connection with entropy of horizons in GR.

At this stage, we have not put any restriction of the boundary $\dV$
or on the choice of the  vector field $\xi^a$. The expression in \eqn{on-shell-2} is
valid for \textit{any} vector field $\xi^a$ ---  not necessarily
null. (Our entropy functional is defined for any vector
field; to obtain the equation of motion we consider only the variation
of null vector fields but having done that, we can study the on-shell entropy for
 any vector field.) The only restriction is that the expression
in \eqn{on-shell-2} should be evaluated on a solution to the field
equations. It is clear that one cannot say much about the value of
this expression in such a general context, keeping the boundary and
$\xi^a$ totally arbitrary. Further, even in the case of a null vector
field $\xi^a$,   the integrand \Cal{I}\ in \eqn{on-shell-2} changes by $\Cal{I}\to f^2(x)\Cal{I}$,
under a rescaling $\xi^a\to f(x)\xi^a$ which keeps the null vector as null. Since the value
of the integral can be changed even by such a rescaling, it is clear
that  a choice has to be made for the overall scaling of the null
vector field    before we can evaluate $S[\xi]$ on-shell
\cite{comment2}. 

The fact that  \eqn{on-shell-2} has no clear interpretation in general should not be surprising since  we  expect to obtain a nontrivial value for the entropy only  in specific cases in which
the solution has a definite thermodynamic interpretation and the surface and the vector field is chosen appropriately.  Obviously, making this connection will
require choosing a particular solution to the field equations, a
particular domain of integration for the entropy functional, etc. and making other specific assumptions.
We shall now calculate the
extremum value in specific situations  and demonstrate that it gives the
standard result for the gravitational entropy when the latter is well-defined and understood.  We will also
discuss several features of this issue and, in particular, will
demonstrate that in the standard cases with horizons, the extremal
value of the entropy correctly reproduces the known results,
\textit{not only in GR but even for \LL theory.} 

The most important case corresponds to solutions with a \textit{stationary} horizon which can be locally approximated as Rindler spacetime. Many of the results which had motivated us to develop the current formalism were proved in this specific context and hence this will act as a natural
testing ground. In this case, the relevant part of the boundary will be a null surface and we will choose
 $\xi^a$ to be a spacelike vector which can be interpreted as describing the displacement of the horizon normal to itself. 
To define this properly, we will  use a limiting procedure and provide
the physical motivation for the choice of $\xi^a$ (based on  certain
locally accelerated observers) thereby leading to a meaningful
interpretation of the on-shell value of $S[\xi]$. 

To set the stage for calculations that follow, we will begin by
briefly recalling the notion of \emph{Rindler} observers in flat
(Minkowski) spacetime. In Minkowski spacetime with inertial coordinates 
$x^i_{\rm M}=(t_{\rm M},x^\alpha_{\rm M})$, $\alpha=1,2,...,D-1$,
observers undergoing constant acceleration along the $x^1_{\rm M}$
direction (the Rindler observers) follow hyperbolic trajectories
\cite{MTW,paddy1} described by $(x^1_{\rm M})^2-(t_{\rm M})^2={\rm
  constant}$. A natural set of coordinates for these observers is
given by $x^a_{\rm R}=(t_{\rm R},N,x^A_\perp)$, $A=2,3,...,D-1$, where
the transformation between $x^a_{\rm R}$ and $x^a_{\rm M}$ are given
by 
\begin{align}
t_{\rm M}=\frac{N}{\kappa}\sinh(\kappa t_{\rm R}) ~~&;~~ x^1_{\rm
  M}=\frac{N}{\kappa}\cosh(\kappa t_{\rm R}) 
\nonumber\\
N=\kappa\left((x^1_{\rm M})^2-(t_{\rm M})^2\right)^{1/2} ~~&;~~
  t_{\rm R}=\frac{1}{\kappa}\tanh^{-1}\left(\frac{t_{\rm
  M}}{x^1_{\rm M}}\right)   
\nonumber\\
x^A_{\rm M}=x^A_\perp ~~&,~~ A=2,3...,D-1\,,
\label{Rind-1}
\end{align}
with constant $\kappa$. The metric in the Rindler coordinates becomes   
\begin{equation}
ds^2=-N^2dt_{\rm R}^2 + dN^2/\kappa^2 + dL_\perp^2\,,
\label{Rind-2}
\end{equation}
where $dL_\perp^2$ is the (flat) metric in the transverse spatial
directions. It is easy to see that the surface described by $(x^1_{\rm
  M})^2-(t_{\rm M})^2=0$ (or $N=0$ in the Rindler coordinates) is
simply the null light cone in the $t_{\rm M}-x^1_{\rm M}$ plane at the
origin, and that it acts as a horizon for the observers maintaining
$N={\rm constant}\neq0$. 

In a general curved spacetime, one can introduce a notion of local
Rindler frames along similar lines. We first go to the local inertial 
frame (LIF, hereafter) around any event \Cal{P}\ and introduce the LIF
coordinates $x^i_{\rm M}=(t_{\rm M},x^\alpha_{\rm M})$,
$\alpha=1,2,...,D-1$. We then use the transformations in \eqn{Rind-1}
to define a local Rindler frame (LRF, hereafter). The choice of $x^1_{\rm M}$
axis is of course arbitrary and one could have chosen any direction in
the LIF as the $x^1_{\rm M}$ axis by a simple rotation. In particular,
a general null surface \Cal{H}\ in the original spacetime passing 
through \Cal{P} can be locally mapped to the null cone in LIF which
--- in turn ---  can be locally identified with the $N=0$ surface for the
\emph{local} Rindler frame. This local patch $\Cal{H}_{\rm
  LIF}\subset\Cal{H}$ of the original null surface acts as a horizon
for these observers. We will make good use of this observation below.
(The local nature of the construction is more
transparent in the Euclidean description. If we choose a LIF around
any event and then transform to an LRF, then the null surface in the
Minkowski coordinates gets mapped to the origin of the Euclidean
coordinates. Our constructions in a local region around the origin in
the Euclidean sector captures the physics near the Rindler horizon in
the Minkowski frame.)  
  The crucial fact to notice
is that locally, \Cal{H}\ is the \emph{Killing} horizon for a suitable 
class of Rindler observers. To see this, choose some point
$\Cal{P}\in\Cal{H}$ and erect a $D$-ad (the $D$-dimensional
generalisation of a tetrad) in the LIF at \Cal{P}, endowed with 
Minkowski coordinates. Let $\Cal{H}_{\rm LIF}\subset\Cal{H}$ denote
the part of \Cal{H}\ contained in the LIF. Choose the $D$-ad in such a
way that the only non-vanishing components of the generator $\chi^a$
of $\Cal{H}_{\rm LIF}$ are $\chi^0$ and $\chi^1$. In other words, with
respect to this $D$-ad, $\Cal{H}_{\rm LIF}$ is defined by $(x^1_{\rm
  M})^2-(x^0_{\rm M})^2=0$, where $x^i_{\rm M}$ are the Minkowski 
coordinates in the LIF. Now transform to the local Rindler frame using
the transformation in \eqn{Rind-1} and consider the vector
$v^a=(1,\vec{0})$ in the Rindler frame. Clearly $v^a$ is the Killing 
vector associated with time translations in the Rindler frame, with
norm $v^av_a=-N^2$, and hence $\Cal{H}_{\rm LIF}$ (given by $N=0$)
is a Killing horizon for the Rindler observers, generated by
$v^a$. (It can be shown that the original generator $\chi^a$ of
$\Cal{H}_{\rm LIF}$ when transformed to the Rindler frame, is
proportional to the Rindler Killing vector $v^a$ \emph{on} the
horizon $\Cal{H}_{\rm LIF}$.)  

 We will now give a
prescription for the evaluation of $S|_{\rm on-shell}$ in a specific
LIF (i.e. on $\Cal{H}_{\rm LIF}\subset\Cal{H}$), which extends to the
entire surface \Cal{H}\ in an obvious way. For notational convenience 
therefore, we will drop the subscript `LIF' on \Cal{H}\ . Instead of the
surface $\Cal{H}$, consider the surfaces in the local Rindler frame at
\Cal{P}\ given by $N=\epsilon=\,$constant. Take $\xi^a=n^a$ as the
unit \emph{spacelike normal} to these surfaces, so that 
\begin{equation} 
n^a=\xi^a=(0,\kappa,0,0,...);\quad\sqrt{h}=\epsilon\sqrt{\sigma}\,,
\label{ent-limit-1}
\end{equation}
where $\sigma$ is the metric determinant on the $t_{\rm R}={\rm
constant}$, $N={\rm constant}$  surfaces. We will evaluate the surface integral for $S|_{\rm on-shell}$  on a surface with
$N=\epsilon=\,$constant, and take the limit $\epsilon\to 0$ at the end of the calculation. The vector displacement field $\xi^a$ then has the natural interpretation of
moving the surface  $N=\epsilon=\,$constant normal to itself and previous work has shown that  these displacements play a crucial role in the thermodynamic interpretation \cite{aseem-sudipta,paddy2}.
In our limiting procedure, we use the normal vector to the surface $N=\epsilon=\,$constant,
fix its norm when the surface is not null (i.e., $\epsilon\neq0$) and
take the $\epsilon\to 0$ limit right at the end so that it remains properly normalised. In this process, we are
considering the null surface as a limit of a sequence of timelike surfaces.
 This is clearly only an ansatz and -- as we have said before -- the
 final result will be different for a different ansatz. In this sense,
 it is the end result that we obtain which 
provides further justification for this choice. 
This prescription can be understood in 
a natural way in the Euclidean continuation ($t_{\rm R}\to it_{\rm
  R}$) of the Rindler frame, where the $N=\epsilon$ surfaces are
circles of radius $\epsilon/\kappa$ in the $t-x$ plane of the
Euclideanised Minkowski coordinates. 

Computing the entropy functional
using this  vector field, and taking the $\epsilon\to0$ limit
at the very end, we find  that     
\begin{equation}
S|_{\Cal{H}} = \sD{4\pi m c_m \int_{\Cal{H}}{d^{D-2}x_{\perp} 
  \sqrt{\sigma}\Cal{L}^{(D-2)}_{(m-1)}}} 
  \,,      
\label{ent-limit-2}
\end{equation} 
where $x_{\perp}$ denotes the transverse coordinates on \Cal{H},
$\sigma$ is the determinant of the intrinsic metric on \Cal{H} and we
have restored a summation over $m$ thereby giving the result for the
most general \LL case. The proof of \eqn{ent-limit-2} can be found in
the Appendix\eqref{sec:entropy}.  The expression in \eqn{ent-limit-2} \emph{is
  precisely the entropy of a general Killing horizon in \LL gravity}
based on the general prescription given by Wald and others
\cite{noether} and computed by several authors \cite{LLentropy}.  This 
result justifies the choice of vector field $\xi^a$ used to compute the
entropy functional (as well as the nomenclature `entropy functional'
itself).  For a wide class
of Killing horizons ($\Cal{H}_{\rm K}$) it is possible to take the \emph{Rindler limit}
of the near-horizon geometry, and write $ds^2=-N^2dt^2 + dN^2/\kappa^2
+ dL_\perp^2$ near the horizon, where $dL_\perp^2$ denotes the line 
element on the transverse surfaces (and in particular on the $N=0$
surface which is the horizon; around any point $\Cal{P}\in\Cal{H}_{\rm K}$
the transverse directions will be locally flat in the Rindler limit). In this case, $\kappa$ is the surface
gravity of the horizon (being constant over the entire horizon)  and we choose  $\xi^a$  by the above limiting prescription. Our construction
is then valid over the entire surface $\Cal{H}_{\rm K}$ and
  the resulting on-shell value of
the entropy functional is precisely the standard entropy of the horizon.

To summarise, we get meaningful results in two cases of importance. First, whenever we have solution to the field equations which possesses a stationary horizon with Rindler limit, we have a natural choice for $\xi^a$ through a limiting procedure such that the extremum value of the entropy functional on shell matches with the standard result for the entropy of horizon. Second, in any spacetime, 
if we take a local Rindler frame around  any event 
we will obtain an entropy for the locally defined Rindler horizon. In the case of GR, this entropy per unit transverse area is just 1/4 as expected. This requires working in a local patch and accepting the notions of local Rindler observers and local Rindler horizons about which there is still no universal agreement. (Not everyone is comfortable with deSitter universe having an observer dependent entropy let alone Rindler horizon, but we believe this is the correct paradigm.)
Finally, in the situation
in which the boundary \dV\ is nowhere null  and
$\xi^a$ is an arbitrary  vector field -- as we pointed out before -- it is hard to say anything about the value of $S$. We hope to investigate this case fully in a later work.

\section{The results in the language of forms}

The purpose of this Section is to recast our formalism in the language
of forms for the sake of those who find such things attractive. This
will, hopefully, help in further work because of two reasons. First,
the expression for Wald entropy \cite{noether} can be expressed in the
language of forms nicely. Second, the action for \LL gravity can be
expressed in terms of the wedge product of curvature forms. We shall
begin with brief pedagogy to set the stage and notation and then will
derive the key results.  

Since the tensor $P^{ab}_{\ph{a}\ph{b}cd}$ has the same algebraic
structure as curvature tensor, one can express it in terms of a
2-form analogous to the curvature 2-form.  We define a 2-form
$\rmb{P}^{ab}$ related to our tensor $P^{ab}_{\ph{a}\ph{b}cd}$ by: 
\begin{equation}
\rmb{P}^{ab}\equiv
\frac{1}{2!}P^{ab}_{\ph{a}\ph{b}cd}\,\om^c\wedge\om^d \,.
\label{form-1}
\end{equation}
Throughout this section we will assume a coordinate 1-form basis
$\om^i=\ext x^i$. If $\rmb{v}=\rmb{e}_av^a$ is a vector, then the
vector-valued 1-form $\ext\rmb{v}$ is given by
\begin{equation}
\ext\rmb{v} = \rmb{e}_a\left(\ext v^a+\om^a_{\ph{a}b}v^b \right) =
\rmb{e}_a\left(\D_bv^a\right)\om^b ~~;~~ \om^a_{\ph{a}b} =
\Gamma{}^a_{bc}\om^c \,.
\label{form-2}
\end{equation}
We will work in the case of pure gravity (that is, $T_{ab}=0$) since
this is more geometrical and since it does not affect the value of the
final on-shell entropy functional. Our entropy functional in
\eqn{ent-func-2} has the integrand ($D$-form): 
\begin{equation}
\Cal{I} = \left(4 P_{ab}^{\ph{a}\ph{b}cd}\D_c\xi^a\D_d\xi^b\right)\msr
\,,
\label{form-3}
\end{equation}
where $\msr=(1/D!)\epsilon_{a_1\cdots a_D}\om^{a_1}\wedge \cdots
\wedge \om^{a_D}$ is the natural $D$-form on the integration domain
\Cal{V}. The first point to note is that the $D$-form in
\eqn{form-3}  is the same as the following: 
\begin{equation}
\Cal{I}= 4 \left( \ast\rmb{P}_{ab} \wedge (\ext\bm{\xi})^a \wedge
(\ext\bm{\xi})^b \right)  \,.
\label{form-4}
\end{equation}
where $(\ext\bm{\xi})^a=(\D_c\xi^a)\om^c$ etc. To see this, we
use 
\begin{equation}
\ast\rmb{P}_{ab} =
\frac{1}{(D-2)!}\frac{1}{2!}P_{ab}^{\ph{a}\ph{b}cd}\epsilon_{cd
  a_1\cdots a_{D-2}}\om^{a_1}\wedge \cdots \wedge\om^{a_{D-2}} \,, 
\label{form-5}
\end{equation}
which allows us to expand the $D$-form in \eqn{form-4} as
\begin{equation}
\Cal{I}= \frac{2}{(D-2)!}P_{ab}^{\ph{a}\ph{b}cd}\epsilon_{cd
  a_1\cdots a_{D-2}} (\D_{a_{D-1}}\xi^a) (\D_{a_D}\xi^b)\om^{a_1}
  \wedge \cdots \wedge \om^{a_D} 
\label{form-6}
\end{equation}
Since this is a $D$-form in a $D$-dimensional space, the right hand
side of   \eqn{form-6} should be expressible as $f\msr$ where $f$ is a  
scalar. This implies
\begin{equation}
\frac{2}{(D-2)!}P_{ab}^{\ph{a}\ph{b}cd}\epsilon_{cd a_1\cdots a_{D-2}}
(\D_{a_{D-1}}\xi^a) (\D_{a_D}\xi^b) = \frac{1}{D!}f\epsilon_{a_1
  \cdots a_D} 
\label{form-7}
\end{equation}
We now use the identity
\begin{equation}
\epsilon^{b_1 \cdots b_{D-j}a_1\cdots a_j} \epsilon_{b_1 \cdots
  b_{D-j}c_1\cdots c_j} = (-1)^s (D-j)!
\Alt{a_1}{a_2}{a_j}{c_1}{c_2}{c_j} \,, 
\label{form-8}
\end{equation}
where $s$ is the number of minus signs in the metric and
$\delta^{\cdots}_{\cdots}$ is the alternating tensor with our 
normalisation.  This also implies 
\begin{equation}
\epsilon^{a_1\cdots a_D}\epsilon_{a_1\cdots a_D}=(-1)^s D! \,.
\label{form-9}
\end{equation}
Contracting both sides of \eqn{form-7} with $\epsilon^{a_1\cdots
  a_D}$ and using the symmetries of $\epsilon_{a_1\cdots a_D}$ leads to
\begin{equation}
f=2 P_{ab}^{\ph{a}\ph{b}cd} (\D_i\xi^a) (\D_j\xi^b)
\delta{}_{cd}^{ij} = 4P_{ab}^{\ph{a}\ph{b}cd} (\D_c\xi^a) (\D_d\xi^b)
\,, 
\label{form-10}
\end{equation}
which is exactly what is required in \eqn{form-3}. Therefore the
entropy functional for any vector field can  be written, somewhat more geometrically, as 
\begin{equation}
S[\xi]=\int_\Cal{V}{4\left( \ast\rmb{P}_{ab}\wedge (\ext\bm{\xi})^a
  \wedge (\ext\bm{\xi})^b\right)} 
\label{form-11}
\end{equation}
We will now do two things. The first is to derive the equations of
motion and hence the on-shell value of $S[\xi]$, and the second is to
relate this value to the Wald entropy. To derive the equations of
motion, it is convenient to introduce some new notation which makes
the derivation more compact. (Essentially, we will suppress all
explicit index occurrences in, say \eqn{form-11} etc.). The relation
to Wald entropy, however, is more easily seen \emph{with} the indices
explicitly in place, so we will revert to the explicit notation at
that stage.  

To get rid of tagging along the indices, we will first introduce the
convention that in a tensor valued $p$-form \rmb{T}, the tensor
indices will always be thought of as being superscripts. For example
if \rmb{T}\ is a 2-tensor valued $p$-form, we have
$\rmb{T}=\rmb{e}_a\rmb{e}_b\rmb{T}^{ab}$ where $\rmb{T}^{ab}$ is a 
$p$-form and we have suppressed the direct product sign for the basis
vectors.  We come across $p$-forms like $\rmb{P}^{ab}$ which are
antisymmetric in $a$ and $b$. These can be denoted by the ``bi-vector
valued'' $p$-form $\rmb{P}=(1/2!)\rmb{e}_a\wedge\rmb{e}_b
\rmb{P}^{ab}$. This will allow us to consistently introduce a dot
product for the tensor valued forms. Let us now introduce the notation
``wedge-dot''  $\wdot$ as follows in terms of couple of examples:  (1)
If \rmb{A}\ is a 2-tensor valued $p$-form and \rmb{B}\ is a vector
valued $q$-form then 
\begin{equation}
\rmb{A}\wdot\rmb{B} \equiv
\rmb{e}_a(\rmb{e}_b\cdot\rmb{e}_c)\rmb{A}^{ab}\wedge\rmb{B}^c 
= \rmb{e}_a\rmb{A}^a_{\ph{a}b}\wedge\rmb{B}^b  \,,
\label{form-12}
\end{equation}
That is, the wedge in $\wdot$ acts in the usual fashion on the $p$-forms
and the dot acts on the two nearest basis vectors it finds. This last
condition also implies that: (2) If \rmb{A}\ is a bivector valued $p$-form 
($\rmb{A}^{ab}=-\rmb{A}^{ba}$) and \rmb{B}\ is a vector valued
$q$-form, then
\begin{equation}
\rmb{A}\wdot\rmb{B} = (-1)^{pq+1}\rmb{B}\wdot\rmb{A}\,,
\label{form-13}
\end{equation}
where the $(-1)^{pq}$ is the usual factor on exchange of the wedge
product, and the additional factor of $(-1)$ arises due to the dot in
$\wdot$ shifting from one index of \rmb{A}\ to the other.

In this notation, we have
\begin{equation}
\ast\rmb{P}_{ab} \wedge (\ext\bm{\xi})^a \wedge (\ext\bm{\xi})^b =
(-1)^{D-2} (\ext\bm{\xi})^a \wedge \ast\rmb{P}_{ab} \wedge
(\ext\bm{\xi})^b = (-1)^{D-2}\,
\ext\bm{\xi}\wdot\ast\rmb{P}\wdot\ext\bm{\xi} \,.
\label{form-14}
\end{equation}
Further, in this notation one can show that for the bivector valued
2-form \rmb{P}, the following relation holds
\begin{equation}
\ast\ext\ast\rmb{P}=\frac{1}{2!}\rmb{e}_a\wedge\rmb{e}_b\,\left(
\D_cP^{abc}_{\ph{a}\ph{b}\ph{c}d} \right)\om^d \,,
\label{form-15}
\end{equation}
and the condition $\D_cP^{abcd}=0$ is equivalent to
$\ext\ast\rmb{P}=0$. This also means $\ext(\ast\rmb{P}\cdot\bm{\xi}) 
=(-1)^{D-2}\,\ast\rmb{P}\wdot\ext\bm{\xi}$ (where the ordinary dot is
defined in the obvious way on the nearest basis vectors), and the
entropy functional becomes 
\begin{align}
S[\xi] &= \int_\Cal{V}{(-1)^{D-2}\,4\bigg[\ext\bm{\xi} \wdot
  \ast\rmb{P} \wdot \ext\bm{\xi}\bigg]} \nonumber\\
  &= \int_\Cal{V}{4\bigg[\ext\bm{\xi} \wdot \ext\left(
  \ast\rmb{P}\cdot\bm{\xi} \right)\bigg] } \,.
\label{form-16}
\end{align}
Using the identity
\begin{equation}
\ext\left( \ext\bm{\xi} \wdot \ast\rmb{P}\cdot\bm{\xi} \right) =
\ext^2\bm{\xi} \wdot \ast\rmb{P}\cdot\bm{\xi} - \ext\bm{\xi} \wdot
\ext\left( \ast\rmb{P}\cdot\bm{\xi} \right)\,,
\label{form-17}
\end{equation}
and the definition of the (bivector valued) Riemann curvature 2-form
via 
\begin{equation}
\ext^2\bm{\xi}= \rmb{R}\cdot\bm{\xi} = -\bm{\xi} \cdot\rmb{R} ~~;~~ 
\rmb{R}=\frac{1}{2!}\rmb{e}_a\wedge\rmb{e}_b \rmb{R}^{ab} ~~;~~
\rmb{R}^{ab} = \frac{1}{2!}R^{ab}_{\ph{a}\ph{b}cd}\om^c \wedge \om^d
\,,  
\label{form-18}
\end{equation}
the entropy functional can be rewritten as
\begin{equation}
S[\xi] = -\int_\Cal{V}{4\left[\bm{\xi}\cdot\rmb{R} \wdot \ast\rmb{P} 
      \cdot \bm{\xi}\right]} - \int_{\dV}{4\left( \ext\bm{\xi} \wdot
      \ast\rmb{P} \cdot\bm{\xi} \right)}\,,
\label{form-19}
\end{equation}
where \dV\ is the $(D-1)$-dimensional boundary of the volume
\Cal{V}. 

We will now specialize to a subset of vector fields which are null and  vary the entropy functional with respect to them. The first term in \eqn{form-19} is manifestly
symmetric in $\bm{\xi}$ and will give a contribution with a factor of
$2$. (This can also be verified using the rules of exchanging the
$\wdot$ etc., or by explicitly working with the indices in place.)
The second term of \eqn{form-19} will not contribute since the 
variation $\delta\bm{\xi}$ vanishes on the boundary. The condition
$\bm{\xi}\cdot\bm{\xi}=0$ is preserved by adding a Lagrange multiplier
term $\lambda\bm{\xi}\cdot\bm{\xi}$ to the entropy functional. The
resulting equations of motion are therefore
\begin{equation}
-4\bm{\xi}\cdot\rmb{R} \wdot \ast\rmb{P} + \lambda\bm{\xi}=0 \,,
\label{form-20}
\end{equation}
which is a set of vector equations. Taking a dot product with
$\bm{\xi}$ we get
\begin{equation}
-4\bm{\xi}\cdot\rmb{R} \wdot \ast\rmb{P}\cdot\bm{\xi} =0 \,,
\label{form-21}
\end{equation}
which must hold for all null vector fields $\bm{\xi}$. \eqn{form-21}
can clearly also be written (since $\ext\ast\rmb{P}=0$) as
\begin{equation}
4\left[\ext\left( \ext\bm{\xi} \wdot \ast\rmb{P}
  \right)\right]\cdot\bm{\xi}=0\,, 
\label{form-22}
\end{equation}
which can also be derived directly by the variation of
\eqn{form-16}. It is easy to show that \eqn{form-21} is equivalent
(after bringing back all indices) to the vacuum version of
\eqn{ent-func-6}. To see this, consider
\begin{align}
-4\bm{\xi}\cdot\rmb{R} \wdot \ast\rmb{P}\cdot\bm{\xi} &=
  -4\xi^a\rmb{R}_{ab} \wedge \ast\rmb{P}^{bc}\xi_c \nonumber\\  
&= -\frac{1}{(D-2)!}\xi^aR_{abk_1k_2}P^{bcmn}\xi_c \epsilon_{mn k_3
  \cdots k_D} \om^{k_1} \wedge \cdots \om^{k_D}\,.
\label{form-22-1}
\end{align}
On equating the right hand side of \eqn{form-22-1} to $f\msr$ where
$f$ is a scalar and using arguments similar to the ones following
equation \eqn{form-6}, we find
\begin{equation}
f=2P_{mb}^{\ph{m}\ph{b}ij}R^m_{\ph{m}aij}\xi^a\xi^b =
2P_b^{\ph{b}ijk}R^a_{\ph{a}ijk}\xi_a\xi^b\,, 
\label{form-22-2}
\end{equation}
which is what appears in the first term on the left hand side of
\eqn{ent-func-6}. The rest of the derivation of field equations
follows as before. 

Having derived the equations of motion by extremising the entropy functional with respect to 
the null vector fields, we go back to the entropy functional for an arbitrary vector field and evaluate it on-shell. Using Eqns. \eqref{form-19} and \eqref{form-21}, the on-shell value of
the entropy functional for an arbitrary vector field is 
\begin{equation}
S|_{\rm on-shell}= -4\int_{\dV}{\left(\ext\bm{\xi} \wdot \ast\rmb{P} 
  \cdot\bm{\xi}\right)} \,.
\label{form-23}
\end{equation}
To show that the on-shell value agrees with Wald entropy for our specific choice described earlier, we take
\dV\ to be a null surface and denote an arbitrary spacelike
cross-section of \dV\ by \Cal{H}. Reverting to (semi)index notation,
we have, 
\begin{align}
S|_{\rm on-shell} &=
-4\int_{\dV}{(\ext\bm{\xi})_a \wedge \ast\rmb{P}^{ab}\xi_b}
\nonumber\\ 
&= -4\int_{\dV}{(\D_k\xi_a)\left(\frac{1}{(D-2)!}\frac{1}{2!}
  P^{abcd}\xi_b \epsilon_{cda_1\cdots a_{D-2}} \right) \om^k \wedge
  \om^{a_1} \wedge\cdots\wedge \om^{a_{D-2}}}\,. 
\label{form-24}
\end{align}
To show that this expression is same as the Wald entropy, we will
again employ the limiting procedure used in section 3; namely, we will
consider the Rindler limit of the geometry near the null surface
\dV.  We consider the vector field
$\bm{\xi}$ to be normal to surfaces of $N={\rm constant}$ for the
Rindler metric \eqn{Rind-2} and take the $N\to0$ limit at the end. The
spacelike cross-section \Cal{H}\ of \dV\  corresponds to the
transverse directions of the Rindler metric and has coordinates
labeled with upper-case indices $x_\perp^A$, $A=2,3,\cdots,D-1$. 

For the vector $\bm{\xi}=\xi^a\rmb{e}_a=(0,\kappa,0, ...)$, we have
$\D_k\xi_a=-(1/\kappa)\Gamma{}^N_{ka}$ of which only the component
$\Gamma{}^N_{00}=\kappa^2N\neq0$. The integrand \Cal{I}\ of
\eqn{form-24} becomes
\begin{equation}
\Cal{I}=4N\left(\frac{1}{(D-2)!}\frac{1}{2!}\right)
P^{0Ncd}\epsilon_{cda_1\cdots a_{D-2}}\om^0 \wedge \om^{a_1} \wedge
\cdots \wedge \om^{a_{D-2}}\,.
\label{form-25}
\end{equation}
Since the integrand is a $(D-1)$-form restricted to a surface of
constant $N$, and since the basis 1-form $\om^0$ appears explicitly,
it follows that the remaining basis 1-forms must be intrinsic to the
spacelike cross-section \Cal{H}. The indices $a_1\cdots a_{D-2}$ in 
\eqn{form-25} can then be replaced by $A_1\cdots A_{D-2}$, and the
integrand reduces to
\begin{equation}
\Cal{I} =
4N\left(g^{00}g^{NN}P_{0N}^{\ph{0}\ph{N}0N}\right)
\sqrt{-g_{00}g_{NN}}\frac{1}{(D-2)!}\sqrt{\sigma}  
e_{A_1\cdots A_{D-2}}\om^0 \wedge \om^{A_1} \wedge \cdots \wedge
\om^{A_{D-2}}  \,,
\label{form-26}
\end{equation}
where $e_{A_1\cdots A_{D-2}}$ is the alternating symbol which takes
values $+1(-1)$ when the indices are an even (odd) permutation of
$2,3,\cdots,D-1$, and is zero otherwise. The metric coefficients refer
to the Rindler metric in \eqn{Rind-2}, and cancel the $N$ dependence of
the integrand. Recognizing the natural $(D-2)$-form on \Cal{H}\ given by 
\begin{equation}
\tilde{\bm{\epsilon}} = \frac{1}{(D-2)!} \sqrt{\sigma} e_{A_1\cdots
  A_{D-2}} \om^{A_1}\wedge \cdots \wedge \om^{A_{D-2}}\,, 
\label{form-27}
\end{equation}
the integrand becomes
\begin{equation}
\Cal{I}=-4\kappa P_{0N}^{\ph{0}\ph{N}0N}\om^0
\wedge\tilde{\bm{\epsilon}}\,.  
\label{form-28}
\end{equation}
In a standard derivation of the expression for the Wald entropy,
e.g.- Ref. \cite{ted}, one would be dealing with the binormal to the
cross-section \Cal{H}, defined in terms of two null vectors normal to
\Cal{H}. Since we are dealing with a limiting procedure in which the
horizon (null surface \dV) is approached in the limit $N\to0$, we 
define a 2-form $\rmb{n}=(1/2!)n_{ij}\om^i\wedge\om^j$ which reduces
to the standard binormal to \Cal{H}\ on the horizon. To do this we use
the fact that in the standard case, the natural $(D-2)$-form
$\tilde{\bm{\epsilon}}$ on \Cal{H}\ is simply the dual of the 2-form
binormal $\rmb{n}$. We therefore define our `binormal' via the relation
\begin{equation}
n^{ij}=\frac{(-1)^s}{(D-2)!} \sqrt{\sigma} e_{A_1\cdots A_{D-2}}
\epsilon^{ijA_1\cdots A_{D-2}}\,.
\label{form-29}
\end{equation}
It is now straightforward to check, using
$\sqrt{-g}=\sqrt{-g_{00}g_{NN}}\sqrt{\sigma}$, that
$P^{abcd}n_{ab}n_{cd} = -4P_{0N}^{\ph{0}\ph{N}0N}$. In Appendix A.2 we
have explicitly shown that this quantity is independent of $N$ in the
\LL case, and hence the $N\to0$ limit is trivial. The on-shell value
of the entropy functional therefore becomes 
\begin{equation}
S|_{\rm on-shell} = \kappa\int_{\dV}{P^{abcd} n_{ab}n_{cd}
  \om^0 \,\wedge\, \tilde{\bm{\epsilon}}}\,.
\label{form-31}
\end{equation}
Since $\om^0=\ext t_{\rm R}$, when the on-shell entropy is evaluated
in a solution which is stationary, the above integral splits into a
time integration and an integral over the arbitrary cross-section
\Cal{H}\ of \dV. Then restricting the time integration range to from
$0$ to $2\pi/\kappa$, we get   
\begin{equation}
S|_{\rm on-shell}= 2\pi\oint_\Cal{H}{P^{abcd}n_{ab}n_{cd}
  \tilde{\bm{\epsilon}}}\,, 
\label{form-32}
\end{equation}
which is precisely the expression for Wald's entropy when we recall
that $P^{abcd}=(\partial\Cal{L}/\partial R_{abcd})$ for the \LL type
theories (up to a sign which depends on the sign convention used for
$\tilde{\bm{\epsilon}}$). Using the definition of $n_{ij}$ and
following the algebra in Appendix A.2, one can also easily recover the
expression in \eqn{ent-limit-2}  thereby proving the equivalence of the two approaches.

\section{Discussion}
 
Since we have described the ideas fairly extensively in the earlier
sections as well as the introduction, we will be brief in this section
and concentrate on the broader picture.  
 
We take the point of view that the gravitational interaction described   
through the metric of a smooth spacetime is an emergent,
long-wavelength phenomenon. Einstein's equations provide the lowest
order description of the dynamics and one would expect higher order
corrections to these equations as we probe the smaller scales.    
It was shown in Ref. \cite{paddy2} that the Einstein equation
$G{}^0_0=8\pi T{}^0_0$ for spherically symmetric spacetimes with
horizons can be rewritten in terms of thermodynamic variables and is  
in fact identical to the first law of thermodynamics $TdS=dE+PdV$. In
Ref. \cite{aseem-sudipta} this result was extended to the 
$E{}^0_0=(1/2)T{}^0_0$ equation for spherically symmetric spacetimes 
in \LL gravity where $E{}^a_b$ was defined in \eqn{LL4}. In fact, the
Lorentz invariance of the theory, taken together with the equation 
$E{}^0_0=(1/2)T{}^0_0$ then leads to the full set of equations
$E{}^a_b=(1/2)T{}^a_b$ governing the dynamics of the metric
$g_{ab}$. The invariance under Lorentz boosts in a local inertial
frame maps to translation along the Rindler time coordinate in the
local Rindler frame. Hence the validity of local thermodynamic
description for \textit{all} Rindler observers allows one to obtain
the full set of equations from the time-time component of the
equations.  
Given the key role played by horizons in all these, it seems natural that we should
have an alternative formulation of the theory in terms of the entropy
associated with the horizons. This is precisely what has been attempted
in this paper. 

The first key result of this paper is an alternative variational principle to obtain
not only the Einstein's theory but also the more general \LL
theory. We have shown that there is a natural procedure for 
obtaining the dynamics of the metric (which is now interpreted as a
macroscopic  variable like the density of a solid) using the
functional $S[\xi]$ given in \eqn{ent-func-2}. This functional can be
defined for any vector field $\xi^a$.  When we restrict attention  
to null vector fields, and demand that the entropy associated with all 
 the null vectors should be an extremum, we obtain a condition on the
 background geometry that is equivalent to the dynamical equations of
 the theory. This provides an alternative route without the usual
 problems which arise in the handling of surface terms etc. when the
 metric is varied. Interestingly, our approach selects out the \LL type of
 theories, which are known to have nice properties regarding the
 integrability of the field equations etc.
 
 As an aside, we want to make a remark on the usual derivations of the field equations in \LL theory. To illustrate the point consider
 the familiar $D=4$ case.  We
know that in 4-D, the second order \LL term $\Cal{L}^{(4)}_2$ (the
Gauss-Bonnet term) is a total divergence, and the higher terms
identically vanish. Usually in the literature, one will ignore the Gauss-Bonnet term, since it is a total divergence, and claim that the equations of motion are identical to Einstein's equations. 
One must note, however, that in a situation wherein the Lagrangian
contains a total derivative term, the conventional action principle is well defined
only when all surviving surface terms are held fixed. It is well known
that in Einstein's GR this can be achieved by adding the
Gibbons-Hawking term  to cancel certain surface
contributions. The case of the Gauss-Bonnet term in 4 dimensions
proves to be trickier in terms of defining consistent boundary
conditions. (See Ref. \cite{olea} for an attempt to address this issue,
and for further references.) So the \LL theory, in the conventional formalism based on varying an action principle, faces certain difficulties associated with the boundary term. Our formalism, of course, reproduces the standard  result and the correct entropy and --- as a bonus --- the above difficulties do not arise in our approach since we do not
vary the metric to get the equations of motion.

The second result in the paper is related to providing an
interpretation of the extremum value of this functional. Here we find
that one needs to make some ansatz which --- though physically
well-motivated and natural --- is logically independent of the
derivation of field equations. 
The ideas we have used are closely connected with previous attempts to
interpret the radial displacements of horizons as the key to obtaining
a thermodynamic interpretation of gravitational theories
\cite{paddy2,aseem-sudipta}. When we evaluate the on-shell value of
the entropy functional associated with a normal to a sequence of
timelike surfaces and take the limit when these timelike surfaces
approach the horizon, we obtain the standard entropy of the
horizon. In the case of GR, the result is intuitively obvious since
the integrand of the entropy functional 
has the structure $n_j(K\xi^j+a^j)$ which is essentially the surface
term in Einstein-Hilbert action from which we know that the correct
entropy can be obtained. Remarkably enough the same prescription works
even in the case \LL gravity for which there is no simple intuitive
interpretation. 

Thus, at the least, we have provided an alternative variational
principle to obtain  not only Einstein gravity but also its closely
related extensions, without varying the metric in the
functional. This, by itself, is worth further study  from three
points of view. First, it is important to understand why it works. In
conventional approaches one interprets the extremum value of action in
terms of the path integral prescription in which alternative histories
are explored by a quantum system; here this should correspond to
fluctuations of the light cone structure in some sense. It is not
clear how to make this notion more precise and useful. Second, the
matter sector is --- as usual --- quite ugly and nongeometrical and
one could even claim that it was added by hand. It is not clear
whether the entropy functional, including the matter term has a
geometric interpretation \cite{comment11}. Finally, the work clearly
endows a special status to the \LL theory as a natural extension of GR
within the thermodynamic paradigm. Several previous results, 
especially Ref. \cite{ayan}, have already pointed in this
direction. Given the rich geometrical structure of the \LL theory
(compared to, for example, theories based on $f(R)$ Lagrangians), it
is worth investigating this issue further. 

These results certainly indicate a deep connection between gravity and
thermodynamics \textit{which goes well beyond Einstein's theory}. The general
\LL theory, which is expected to partially account for an effective
action for gravity in the semiclassical regime, satisfies the same
relations between the dynamics of horizons and thermodynamics, as
Einstein's GR. This suggests that these results have possible
consequences concerning a quantum theory of gravity as well. In a
previous work \cite{ayan} it was shown that this class of theories
exhibits a type of classical `holography' which assumes special
significance in the backdrop of current results. 

\acknowledgments
\noindent We thank R. Bousso and R. Nityananda for useful comments. AP
thanks all the members of IUCAA for their warm 
hospitality during his stay there, during which a part of this work
was completed.

\appendix
\section{}\label{sec:mainapp}
\noindent In this appendix we provide proofs for some of the results
quoted in the text.

\subsection{Proof of \eqn{ent-func-7}}\label{sec:loc-Lor}
\noindent It is sufficient to prove that if a symmetric tensor
$S_{ab}$ satisfies 
\begin{equation}
S_{ab}\xi^a\xi^b=0 \,,
\label{app-ll-1}
\end{equation}
for an arbitrary null vector field $\xi^a$, then it must satisfy
$S_{ab}=F(x)g_{ab}$ where $F$ is some scalar function. Consider a
point \Cal{P}\ in the spacetime and construct the local inertial frame
(LIF) at \Cal{P}, endowed with Minkowski coordinates. The vector field
$\xi^a$ appearing in \eqn{app-ll-1} is arbitrary, and the choice of the
$D$-ad erected in the LIF is also arbitrary up to a local Lorentz
transformation (LLT) (which has $D(D-1)/2$ degrees of freedom). We
will utilize the $(D-1)(D-2)/2$ (spatial) rotational degrees of
freedom available in the LLT to choose the $D$-ad axes such that the
purely spatial components of $S_{ab}$ with respect to these axes,
vanish; namely $S_{\alpha\beta}=0$,
$\alpha,\beta=1,2,...,D-1$. Similarly, we utilize the $(D-1)$ boost 
degrees of freedom in the LLT to choose the $D$-ad axes such that
$S_{0\alpha}=0$, $\alpha=1,2,...,D-1$. Further, since the null vector
field $\xi$ is arbitrary, we will in turn consider the vector fields
given by  
\begin{equation}
\xi^a_{(\alpha)}=\delta^a_0+\delta^a_\alpha \,,\alpha=1,2,...,D-1, 
\label{app-ll-2}
\end{equation}
where the components are understood to be with respect to the local
Minkowski coordinates. Substituting for this choice of $\xi^a$ in
\eqn{app-ll-1}, we find 
\begin{equation}
S_{00}+2S_{0\alpha}+S_{\alpha\alpha}=0\,.
\label{app-ll-3}
\end{equation}
The middle term drops out because of our choice of $D$-ad
axes. Repeated application of \eqn{app-ll-3} for all allowed values of 
$\alpha$ then gives us $S_{\alpha\alpha}=-S_{00}$ for all $\alpha$,
and combined with $S_{0\alpha}=0=S_{\alpha\beta}$, we obtain
\begin{equation}
S_{ab}(\Cal{P})=F(\Cal{P})\eta_{ab}\,,
\label{app-ll-4}
\end{equation}
where $F$ is a scalar depending on the choice of \Cal{P}. Since
\eqn{app-ll-4} is a tensor equation in the LIF, it immediately
generalises as required, to
\begin{equation}
S_{ab}(x)=F(x)g_{ab}(x)\,,
\label{app-ll-5}
\end{equation}

\subsection{Proof of \eqn{ent-limit-2}}\label{sec:entropy}
\noindent
 We now consider the algebraic details in the evaluation of the on-shell value of entropy functional. As explained in the text, we expect to obtain meaningful result only for certain solutions, when a specific choice is made for the for the boundary and the vector field. To do this we define it through a limiting process involving a sequence of timelike surfaces and their normals  with the limit taken at the end of the calculation.
 According to the prescription laid down in the text,
we  take 
$\xi^a=n^a$ to be the unit spacelike normal to the $N=\epsilon$
surfaces in the Rindler frame, and taking the $\epsilon\to0$ limit at 
the very end of the calculation.  (This limiting procedure  is physically well motivated; in the case of standard GR and a Schwarzschild black hole, for example, it will correspond to approaching the $r=2M$ surface as the $\epsilon\to0$ limit of $r=2M+\epsilon$ sequence of surfaces.)
In the Rindler frame, with the metric
$ds^2=-N^2dt_{\rm R}^2 +dN^2/\kappa^2 + dL_\perp^2$,  
\begin{equation} 
n_a=\xi_a=(0,1/\kappa,0,0,...) ~~;~~ n^a=\xi^a=(0,\kappa,0,0,...)
~~;~~ \sqrt{h}=\epsilon\sqrt{\sigma}\,,
\label{app-ent-1}
\end{equation}
where $h$ is the metric determinant for the $N=\epsilon$ surfaces and 
$\sigma$ the metric determinant for the $N={\rm constant}$, $t_{\rm 
  R}={\rm constant} $ surfaces. (When $\epsilon\neq0$,  the surface and its normal are not null but $\epsilon=0$ is the null Rindler horizon). The entropy functional is
\begin{equation}
S[\xi]=\int_{\dV,\epsilon}{d^{D-1}x\sqrt{h}\,n_a\left(
  4P^{abcd}\xi_c\D_b\xi_d\right)}\,.
\label{app-ent-2}
\end{equation}
Here, $d^{D-1}x=dt_{\rm R}d^{D-2}x_\perp$, and $\D_b\xi_d =
-\Gamma^a_{\ph{a}bd}\xi_a = -(1/\kappa)\Gamma^N_{\ph{a}bd}$, of which
only $\Gamma^N_{\ph{a}00}=\kappa^2\epsilon\neq 0$. The $\epsilon$ on
the integration symbol reminds us that we are not actually on the
given boundary \dV, but will approach it as $\epsilon\to0$. As
mentioned in the text, our choice of the tensor $P^{abcd}$ is a single
tensor ${}^{(m)}P^{abcd}=m{}^{(m)}Q^{abcd}$, and appears 
linearly in the expression for $S|_{\rm on-shell}$. It is sufficient
to analyse this expression using the single ${}^{(m)}P^{abcd}$ and to
take the appropriate linear combination for the general \LL case at
the end. This will not interfere with the process of taking the limit
$\epsilon\to0$. As in the text, we will drop the superscript $(m)$ for 
notational convenience. The integrand for a single $m$ can be
evaluated as follows  
\begin{align}
\sqrt{h}\,n_a\left(4P^{abcd}\xi_c\D_b\xi_d\right) &=
\frac{\epsilon\sqrt{\sigma}}{\kappa^2}\left(4P^{NbNd}\D_b\xi_d\right)
\nonumber\\ 
&=
\frac{\epsilon\sqrt{\sigma}}{\kappa^2}\left(
-4P^{N0N0}\frac{1}{\kappa}\Gamma^N_{\ph{a}00}\right) \nonumber\\
&= \frac{\epsilon^2\sqrt{\sigma}}{\kappa}\left(-4P^{N0N0}\right)
\nonumber\\ 
&= \frac{\epsilon^2\sqrt{\sigma}}{\kappa}\left(-4 m g^{00}g^{NN}
Q_{N0}^{\ph{N}\ph{0}N0}\right) \nonumber\\
&= \kappa\sqrt{\sigma}\left(4 m Q_{N0}^{\ph{N}\ph{0}N0}\right)\,.
\label{app-ent-3}
\end{align}
Consider the quantity $Q_{N0}^{\ph{N}\ph{0}N0}$ which --- for the $m$-th
order \LL action ---  is given by 
\begin{align}
Q_{N0}^{\ph{N}\ph{0}N0} &= \frac{1}{16\pi}\frac{1}{2^m}
\AltC{N}{0}{a_3}{a_{2m}}{N}{0}{b_3}{b_{2m}}
\left(\Riem{b_3}{b_4}{a_3}{a_4}
... \Riem{b_{2m-1}}{b_{2m}}{a_{2m-1}}{a_{2m}}\right)_{N=\,\epsilon}
\,. 
\label{app-ent-4}
\end{align}
The presence of both $0$ and $N$ in each row of the alternating tensor
forces all other indices to take the values $2,3,...,D-1$. In fact, we
have $\AltC{N}{0}{a_3}{a_{2m}}{N}{0}{b_3}{b_{2m}} =
\Alt{A_3}{A_4}{A_{2m}}{B_3}{B_4}{B_{2m}}$ with $A_i, B_i =
2,3,...,D-1$ (the remaining combinations of Kronecker deltas on
expanding out the alternating tensor are all zero since
$\delta^0_A=0=\delta^N_A$ and so on). Hence $Q_{N0}^{\ph{N}\ph{0}N0}$
reduces to 
\begin{align}
Q_{N0}^{\ph{N}\ph{0}N0} &=
\frac{1}{2}\left(\frac{1}{16\pi}\frac{1}{2^{m-1}}\right) 
\Alt{A_3}{A_4}{A_{2m}}{B_3}{B_4}{B_{2m}}
\left(\Riem{B_3}{B_4}{A_3}{A_4}
... \Riem{B_{2m-1}}{B_{2m}}{A_{2m-1}}{A_{2m}}\right)_{N=\,\epsilon}
\,. 
\label{app-ent-5}
\end{align}
In the $\epsilon\to 0$ limit therefore, recalling that
$\Riem{A}{B}{C}{D}\mid_\Cal{H}
=\,^{(D-2)}\Riem{A}{B}{C}{D}\mid_\Cal{H}$, we find that $Q_{N0}^{\ph{N}\ph{0}N0}$ is essentially the \LL Lagrangian of order $(m-1)$:
\begin{equation}
Q_{N0}^{\ph{N}\ph{0}N0}=
\frac{1}{2}\Cal{L}^{(D-2)}_{(m-1)}\mid_\Cal{H} \,,
\label{app-ent-6}
\end{equation} 
and the entropy functional becomes
\begin{equation}
S|_{\Cal{H}} =2m\kappa\int_\Cal{H}{dt_{\rm
  R}d^{D-2}x_\perp\sqrt{\sigma}  
  \Cal{L}^{(D-2)}_{(m-1)}}  \,.
\label{app-ent-7}
\end{equation}
Restricting the $t_{\rm R}$ integral to the range $(0,2\pi/\kappa)$ as usual and using stationarity
(which cancels the $\kappa$ dependence of the result), we get
\begin{equation}
S^{(m)}|_{\Cal{H}} = 4\pi m
\int_{\Cal{H}}{d^{D-2}x_{\perp}\sqrt{\sigma}\Cal{L}^{(D-2)}_{(m-1)}}
\,,   
\label{app-ent-8}
\end{equation}
where we have restored the superscript $(m)$ in the last
expression. Finally, taking the appropriate linear combination we find 
\begin{equation}
S|_{\Cal{H}} = \sD{c_mS^{(m)}|_{\Cal{H}}} =\sD{4\pi m c_m
\int_{\Cal{H}}{d^{D-2}x_{\perp}\sqrt{\sigma}\Cal{L}^{(D-2)}_{(m-1)}}}
\,,   
\label{app-ent-9}
\end{equation}
This is precisely the entropy in the \LL theory. We have also verified
that this expression is explicitly recovered when working with a
spherically symmetric metric of the form
$ds^2=-f(r)dt^2+dr^2/f(r)+r^2d\Omega_{(D-2)}^2$ which admits a horizon
at some value of $r=r_{\Cal{H}}$. The calculation in this case can be
done using this metric (instead of the Rindler metric) and considering
the limit $r\to r_{\Cal{H}}$.

 Note that, instead of the above limiting procedure, if we had just
 foliated the spacetime with null surfaces, chosen $\xi_a$ to be the null normal vector field to the foliating surfaces
  and taken the boundary to be one of the foliating surfaces 
  (so that the normal vector
 $n_a$ coincides with the null vector on the boundary) then the
 surface term  $S|_{\rm on-shell}$ 
will give zero on $\Cal{H}$. This is transparent in Einstein
gravity, where the integrand of $S[\xi]$ becomes $\sim
n_a\left(-\xi^a(\D_b\xi^b) + \xi^b\D_b\xi^a\right)$, which vanishes on
$\Cal{H}$ where $n^a$ and $\xi^a$ coincide and are null. It can be
 shown that the same result holds more generally and is only dependent
 on the algebraic symmetries of $P^{abcd}$. Similarly, if we choose
 $\xi^a=v^a=(1,\vec{0})$ in the local Rindler frame --- which is the
 Rindler time translation Killing vector  that becomes null on the
 horizon --- and use the same limiting procedure, we get a vanishing
 entropy.  On the other hand, the \textit{normalised} vector in the timelike direction $N^{-1}v^a$ gives the same (correct) result as
 as our choice. Clearly the final result depends on our choice and no
 general statement can be made. 

Finally, to make contact with results in a more familiar setting, we
point out that some of these features are not unique 
to the above context. In fact, a similar caveat also applies to the well
known Gibbons-Hawking term which is a similar surface term
arising in the Einstein-Hilbert Lagrangian. Apart from some constant
proportionality factors which are irrelevant to this discussion, this
term is actually arises as the integral of a total derivative (in 4
dimensions) as
\begin{equation}
A^{\rm GH}\sim \int_\Cal{V}{d^4x\sqrt{-g}\D_a\left(n^a\D_bn^b\right)}
= -\int_\Cal{V}{d^4x\sqrt{-g}\D_a\left(n^aK\right)}\sim \int_{\dV}{d^3x\sqrt{h}(v_an^a)K} 
\label{app-GH-1}
\end{equation}
where $n^a$ is the normal to the foliating surfaces, $v_a$ is the normal to the boundary  \dV\ of the 4-dimensional
region \Cal{V}  and $K=-\D_bn^b$ is the trace of the extrinsic
curvature. We now consider the case in which the boundary surfaces are chosen to be members of the set of foliating surfaces (e.g,  we can foliate the spacetime by $t=$ constant surfaces and choose part of \dV\ to be given by $t=t_1$ and $t=t_2$) so that $v_a=n_a$. Then we have
\begin{equation}
A^{\rm GH}\sim \int_{\dV}{d^3x\sqrt{h}(n_an^a)K} \sim
\int_{\dV}{d^3x\sqrt{h}K}\,, 
\label{app-GH-2}
\end{equation}
\emph{provided} \dV\ is \emph{not}
null, so that the normal vector can be assumed to have unit norm $n_an^a=\pm1$.
This is the familiar expression often quoted in the
literature. Clearly, the above naive argument breaks down (but the result still holds) when the spacetime is foliated by a series of null surfaces ($n_an^a=0$) and the boundary is one of these surfaces. But this case also can be handled by a  limiting procedure similar to the one we used for
computing  our surface integral. In fact, our prescription essentially foliates the Rindler limit of the horizon by a series of time like surfaces (like the $r=2M+\epsilon=$constant surfaces in the Schwarzschild) approaching the null horizon in a particular limit (like $\epsilon\to0$).
In the case of GR, this is equivalent to the standard calculation of integrating the extrinsic curvature (defined by this foliation) over the surface and -- of course -- we get the standard result of entropy density being a quarter of transverse area. What is more interesting and nontrivial is that the same prescription works in a much wider context and reproduces the \LL entropy.


\begin{thebibliography}{20} 

\bibitem{elastic} 
A D Sakharov,  {\it Sov. Phys. Dokl.}, {\bf 12}, 1040, (1968); 
T Jacobson,  {\it Phys. Rev. Lett.}, {\bf 75},  1260, (1995); 
T Padmanabhan, {\it Mod. Phys. Lett.}, {\bf A 17}, 1147, (2002) [hep-th/0205278]; 
                                         {\bf 18}, 2903, (2003) [hep-th/0302068]; 
G E Volovik, {\it Phys. Rept.}, {\bf 351}, 195, (2001);
G E Volovik, {\it The universe in a helium droplet}, (Oxford
University Press, 2003);  
B L Hu, [gr-qc/0503067] and references therein.


\bibitem{zeropoint}
H S Snyder, \textit{ Phys. Rev.}, \textbf{71}, 38 (1947);
B S DeWitt,\textit{ Phys. Rev. Lett.}, \textbf{13}, 114 (1964);
T Yoneya  \textit{Prog. Theor. Phys.}, \textbf{56}, 1310 (1976);
T Padmanabhan, \textit{ Ann. Phys.} (N.Y.), \textbf{165}, 38 (1985); 
            \textit{Class. Quantum Grav.}, \textbf{4}, L107 (1987);
A Ashtekar et al., \textit{Phys. Rev. Lett.}, \textbf{69}, 237 (1992 );
T Padmanabhan,  \textit{Phys. Rev. Lett.}, \textbf{78},  1854 (1997) [hep-th/9608182];
              \textit{Phys. Rev.}  \textbf{D 57}, 6206 (1998); 
For a review, see L J Garay, \textit{ Int. J. Mod. Phys.}, \textbf{A10}, 145 (1995).

\bibitem{magglass}
 See e.g., 
 T Padmanabhan, {\it Phys. Rev. Lett.}, {\bf 81}, 4297,(1998) [hep-th/9801015]; 
                 {\it Phys. Rev.} {\bf D59}, 124012, (1999)[hep-th/9801138]
  and references therein. 

\bibitem{stdrefST} 
S A Fulling, \textit{Phys. Rev.} \textbf{D7}, 2850 (1973);
S Hawking, {\it Commun. Math. Phys.}, \textbf{43}:199-220, (1975);
W G Unruh, \textit{ Phys. Rev.} \textbf{D14}, 870 (1976);
G W Gibbons, S W Hawking,  {\it Phys. Rev.} \textbf{D15} (1977) 2752--2756;
K Srinivasan et al.,\textit{ Phys. Rev.} \textbf{D 60}, 24007 (1999) [gr-qc-9812028];
L Sriramkumar et al., \textit{Int. Jour. Mod. Phys.},  \textbf{ D
  11},1 (2002) [gr-qc/9903054] 

\bibitem{paddy1}
For a recent review, see:
T Padmanabhan, {\it Phys. Rept.}, {\bf 406}, 49, (2005),[gr-qc/0311036]; 
               {\it Mod. Phys. Lett.} {\bf A17}, 923, (2002) [gr-qc/0202078]. 

\bibitem{paddyholo}
T Padmanabhan, {\it Brazilian Jour.Phys.} (Special Issue) {\bf 35},
                362, (2005) [gr-qc/0412068];  
               {\it Gen. Rel. Grav.}, {\bf 34}, 2029, (2002) [gr-qc/0205090]; 
               {\it Gen. Rel. Grav.}, {\bf 35}, 2097, (2003).


\bibitem{paddy2}
T Padmanabhan,{\it Class. Quan. Grav.}, {\bf 19}, 5387, (2002) [gr-qc/0204019]. 

\bibitem{aseem-sudipta}
A Paranjape, S Sarkar and T Padmanabhan, {\it Phys. Rev.} {\bf D74},
                          104015, (2006) [hep-th/0607240].  

\bibitem{rongencai}
R-G Cai, L-M Cao, [gr-qc/0611071]; 
M Akbar, R-G Cai, [hep-th/0609128].
 
\bibitem{dawood-sudipta-tp} 
Dawood Kothawala, Sudipta Sarkar, T Padmanabhan, \textit{Einstein's
  equations as a thermodynamic identity: The cases of stationary
  axis-symmetric horizons and evolving spherically symmetric
  horizons}, gr-qc/0701002. 

\bibitem{cai2} 
M. Akbar, Rong-Gen Cai, gr-qc/0612089; 
C. Eling et.al, \textit{ Phys.Rev.Lett.}\textbf{ 96}, 121301, 2006 [gr-qc/0602001].
  
  

\bibitem{obsdepentropy} See e.g.,
T Jacobson and R Parentani, {\it Found. Phys.}, \textbf{33} (2003) 323-348, [gr-qc/0302099]; 
T Padmanabhan,  {\it Class. Quan. Grav.}, \textbf{21}, 4485 (2004) [gr-qc/0308070].

\bibitem{comment0}
These are the two reasons why we do not want to think of the standard
action functional itself as some kind of an entropy, but are looking
for alternatives. At a somewhat ill-defined level, one can think of
Euclidean action as analogous to entropy, but this has no relationship
with null surfaces and blocking of information. Also, in using the
action functional, one considers the metric as dynamical and the 
extremisation is done with respect to variations in the metric. We
will in contrast have a completely different and more rigorous choice
for the entropy functional. 


\bibitem{lan-lif} 
L D Landau and E M Lifshitz, {\it Theory of Elasticity}, 3rd ed.,
Butterworth-Heinemann, (Oxford, UK), (1986).



\bibitem{comment1} 
T Padmanabhan,  {\it Int. Jour. Mod. Phys.} {\bf  D13}, 2293-2298(2004)
[gr-qc/0408051];   [gr-qc/0609012].   
This idea was earlier developed for general relativity in these papers
  and  more detailed conceptual comparison with elasticity can be
  found in these papers.



\bibitem{MTW}
C W Misner, K S Thorne and J A Wheeler, {\it Gravitation}, (W H
Freeman, San Francisco, 1973)   


\bibitem{ayan}
A Mukhopadhyay and T Padmanabhan, {\it Phys. Rev.} {\bf D74},124023, (2006) [hep-th/0608120].

\bibitem{lovelock} 
C Lanczos, {\it Z. Phys.} {\bf 73}, 147, (1932); {\it Annals Math.} {\bf 39}, 842, (1938); 
D Lovelock, {\it J. Math. Phys.}, {\bf 12}, 498 (1971). 

\bibitem{comment11} The following curious fact is worth mentioning. If we treat $R^{ab}_{cd}$ and $g^{ab}$ as independent variables,
then the gravitational Lagrangian \textit{density} in \LL theory is formally independent of $g^{ab}$ and is constructed from
$R^{ab}_{cd}$ and the alternating tensor. On the other hand, the matter Lagrangian density is dependent on $g^{ab}$ but is usually independent of $R^{ab}_{cd}$. If $\mathcal{L}=\mathcal{L}_{grav}+\mathcal{L}_{matter}$ is the total Lagrangian density, then our integrand in the entropy functional is essentially the sum
$4(\partial \mathcal{L}/\partial R^{ab}_{cd})\D_c\xi^a\D_d\xi^b+(1/2)(\partial \mathcal{L}/\partial g^{ab})\xi^a\xi^b$ for any \textit{null} vector field $\xi^a$, because the term proportional to $g_{ab}$ in the definition stress tensor does not contribute. We do not use this result in this paper.





\bibitem{cc1}
T. Padmanabhan, {\it Gen. Rel. Grav.}, {\bf 38}, 1547, (2006);
T Padmanabhan,  {\it Int. J. Mod. Phys.}, {\bf D15}, 1659, (2006) [gr-qc/0606061];  
              {\it Class. Quant. Grav.}, {\bf 22}, L107-L110, (2005) [hep-th/0406060].
T. Padmanabhan, \textit{Dark Energy: Mystery of the
 Millennium}, section 5;  Lecture  at Albert Einstein Century
International Conference, Paris, France, 18-23 July, 2005 [astro-ph/0603114];   




\bibitem{olea}
R Aros, \emph{et al.}, {\it Phys. Rev. Lett.}, {\bf 84}, 1647-1650,
(2000) [gr-qc/9909015]; 
R Aros, \emph{et al.}, {\it Phys. Rev.} {\bf D62}, 044002, (2000)
[hep-th/9912045]; 
R Olea, {\it JHEP} 0506:023, (2005) [hep-th/0504233].



\bibitem{othernull} For a formulation of GR using null surfaces, see 
S Frittelli et al., \textit{J. Math. Phys.}, \textbf{36} (1995) 4984;
\textit{ibid}., 4975; \textit{ibid}., 5005; for a different
perspective on null surfaces, related to holography, see D Cremades
et al., hep-th/0608174. 


\bibitem{comment2} To avoid a possible confusion we make the following
  comment: We mentioned  earlier that \eqn{ent-func-6}
  can be obtained by requiring that $S[\xi]$ be invariant under scaling
transformations of $\xi^a$ while \eqn{on-shell-2} on the other hand shows
that the on-shell value of $S[\xi]$ is \emph{not} invariant under such
a rescaling. These results are not contradictory because the earlier
  result required that the infinitesimal rescaling function 
$\epsilon(x)$ vanish on \dV\ while here we are concerned with this
  \textit{surface term} evaluated on \dV. 

\bibitem{noether}
 R M Wald, {\it Phys. Rev.} \textbf{D48}, 3427 (1993);
 V Iyer and R M Wald, {\it Phys. Rev.}  \textbf{D52}, 4430-4439 (1995)


 


\bibitem{LLentropy} 
There is an extensive literature on the topic of entropy in the
context of higher derivative theories of gravity. For a sample, see
T Jacobson and R C Myers, {\it Phys. Rev. Lett.}, {\bf 70}, 3684, (1993);
R C Myers and J Z Simon, {\it Phys. Rev.} {\bf D38}, 2434, (1998);
R-G Cai, {\it Phys. Rev.} {\bf D65}, 084014, (2002), {\it Phys. Rev. Lett.}, {\bf 582}, 237, (2004);  
S Nojiri, S D Odintsov and S Ogushi, {\it Phys. Rev.} {\bf D65}, 023521, (2002) [hep-th/0108172];
S Nojiri and S D Odintsov, {\it Phys. Lett.} {\bf B521}, 87, (2001) [hep-th/0109122]; 
M Cvetic, S Nojiri and S D Odintsov, {\it Nucl. Phys.} {\bf B628}, 295, (2002) [hep-th/0112045]; 
T Clunan, S F Ross and D J Smith, {\it Class. Quant. Grav.}, {\bf 21}, 3447-3458, (2004); 
I P Neupane, {\it Phys. Rev.} {\bf D67}, 061501, (2003); 
Y M Cho and I P Neupane, {\it Phys. Rev.}  {\bf D66}, 024044, (2002) [hep-th/0202140]; 
N Deruelle, J Katz and S Ogushi, {\it Class. Quant. Grav.}, {\bf 21}, 1971, (2004) [gr-qc/0310098]; 
G Kofinas and R Olea, [hep-th/0606253], (2006). 
For related work see 
R-G Cai and S P Kim, {\it JHEP} 0502:050, (2005) [hep-th/0501055];
M Akbar and R-G Cai, {\it Phys. Lett.}  {\bf B635}, 7, (2006) [hep-th/0602156]. 



\bibitem{ted}
T Jacobson, G Kang and R C Myers, {\it Phys. Rev.} \textbf{D49},
6587-6598, (1994) [gr-qc/9312023].


\end{thebibliography}
\end{document}